\pdfoutput=1

\documentclass[11pt,twoside,a4paper,cmspaper,final,collab]{cms-tdr}
\def\svnVersion{285305}\def\cmsCernNo{2013-037}\def\cmsCernDate{\today}\def\cmsMessage{Published in the European Physical Journal C as \href{http://dx.doi.org/10.1140/epjc/s10052-015-3364-2}{\doi{10.1140/epjc/s10052-015-3364-2.}}}

\begin{document}\cmsNoteHeader{SMP-14-003}

\hyphenation{had-ron-i-za-tion}
\hyphenation{cal-or-i-me-ter}
\hyphenation{de-vices}
\RCS$Revision: 278936 $
\RCS$HeadURL: svn+ssh://svn.cern.ch/reps/tdr2/papers/SMP-14-003/trunk/SMP-14-003.tex $
\RCS$Id: SMP-14-003.tex 278936 2015-02-27 19:44:24Z alverson $
\newlength\cmsFigWidth
\ifthenelse{\boolean{cms@external}}{\setlength\cmsFigWidth{0.98\columnwidth}}{\setlength\cmsFigWidth{0.9\textwidth}}
\ifthenelse{\boolean{cms@external}}{\providecommand{\cmsLeft}{top}}{\providecommand{\cmsLeft}{left}}
\ifthenelse{\boolean{cms@external}}{\providecommand{\cmsRight}{bottom}}{\providecommand{\cmsRight}{right}}
\providecommand{\expstat}{\ensuremath{\,\text{(exp)}}\xspace}
\providecommand{\PP}{\ensuremath{\Pp\Pp}\xspace}
\cmsNoteHeader{SMP-14-003} 
\title{Measurements of differential and double-differential Drell--Yan cross sections in proton-proton collisions at $\sqrt{s} = 8$\TeV}
\titlerunning{Drell--Yan cross sections at $\sqrt{s}=8\TeV$}

\date{\today}

\abstract{
Measurements of the differential and double-differential Drell--Yan cross sections in the dielectron and dimuon channels
are presented. They are based on proton-proton collision data at $\sqrt{s} = 8\TeV$ recorded with the CMS detector at the LHC and corresponding
to an integrated luminosity of 19.7\fbinv.
The measured inclusive cross section in the \cPZ\ peak region (60--120\GeV), obtained from the combination of the
dielectron and dimuon channels, is
$1138 \pm 8\expstat \pm 25\thy \pm  30\lum\unit{pb}$,
where the statistical uncertainty is negligible.
The differential cross section $\rd\sigma/\rd{}m$ in the dilepton mass range 15 to 2000\GeV is measured and corrected to the full phase space.
The double-differential cross section $\rd^2\sigma/\rd{}m\,\rd\abs{y}$ is also measured
over the mass range 20 to 1500\GeV and absolute dilepton rapidity
from 0 to 2.4. In addition, the ratios of the normalized differential cross sections measured at $\sqrt{s} = 7$ and 8\TeV are presented.
These measurements are compared to the predictions of perturbative
QCD at next-to-leading and next-to-next-to-leading (NNLO) orders using various
sets of parton distribution functions (PDFs).
The results agree with the NNLO theoretical predictions computed with \FEWZ 3.1 using the CT10 NNLO and NNPDF2.1 NNLO PDFs.
The measured double-differential cross section and ratio of normalized differential cross sections are sufficiently precise to constrain the
proton PDFs.
}

\hypersetup{%
pdfauthor={CMS Collaboration},
pdftitle={Measurements of differential and double-differential Drell--Yan cross sections in proton-proton collisions at 8 TeV},
pdfsubject={CMS},%
pdfkeywords={CMS, physics, standard model Physics, Drell--Yan}}

\maketitle 

\section{Introduction}

At hadron colliders, Drell--Yan (DY) lepton
pairs are produced via $\gamma^{*}/\cPZ$ exchange in the
$s$ channel. Theoretical calculations of the
differential cross section $\rd\sigma/\rd{}m$ and the double-differential cross section $\rd^2\sigma/\rd{}m\,\rd\abs{y}$,
where $m$ is the dilepton invariant mass and $\abs{y}$ is the absolute value of the dilepton rapidity, are well established
in the standard model (SM) up to the next-to-next-to-leading order (NNLO) in perturbative quantum chromodynamics (QCD)~\cite{QCDNNLO, DYNNLO, DYNNLO1, DY-Theory}.
The rapidity distributions of the gauge bosons $\gamma^{*}/\cPZ$ are sensitive to the parton content of the proton.

The rapidity and the invariant mass of the dilepton system
produced in proton-proton collisions are related at leading order to the
longitudinal momentum fractions $x_+$ and $x_-$ carried by the two interacting
partons according to the formula \( x_\pm=(m/\sqrt{s}) \re^{\pm y}\).
Hence, the rapidity and mass distributions are sensitive to the parton distribution functions (PDFs) of the interacting partons.
The differential cross sections are measured with respect to $\abs{y}$ since the rapidity distribution is symmetric about zero.
The high center-of-mass energy at the CERN LHC permits the study of DY production in regions of the Bjorken scaling variable and evolution scale $Q^2=x_+x_-s$ that were not accessible in previous experiments~\cite{bib:ZEUS,bib:SLAC,bib:FNAL1,bib:FNAL3,bib:FNAL_late1,bib:FNAL_late2}. The present analysis covers the ranges $0.0003 < x_\pm < 1.0$ and $600 < Q^2 < 750\,000\GeV^2$ in the double-differential cross section measurement. The differential cross section $\rd\sigma/\rd{}m$ is measured in an even wider range $300 < Q^2 < 3\,000\,000\GeV^2$.

The increase in the center-of-mass energy at the LHC from 7 to 8\TeV provides an opportunity to measure the
ratios and double-differential ratios of cross sections of various hard processes, including the DY process.
Measurements of the DY process in proton-proton collisions depend
on various theoretical parameters such as the QCD running coupling
constant, PDFs, and renormalization and
factorization scales.
The theoretical systematic uncertainties in the cross section measurements
for a given process at different center-of-mass energies are substantial but correlated, so that
the ratios of differential cross sections normalized to the
\cPZ\ boson production cross section (double ratios) can be measured very precisely~\cite{bib:DDRojo}.

This paper presents measurements of the DY differential cross section $\rd\sigma/\rd{}m$
in the mass range $15 < m < 2000$\GeV, extending the measurement reported in~\cite{Paper7TeV},
and of the double-differential cross section $\rd^2\sigma/\rd{}m\,\rd\abs{y}$
in the mass range $20 < m < 1500$\GeV and absolute dilepton rapidity from 0 to 2.4.
In addition, the double ratios measured at 7 and 8\TeV are presented.
The measurements are
based on a data sample of proton-proton collisions with a center-of-mass energy $\sqrt{s} = 8\TeV$, collected with the CMS detector
and corresponding to an integrated luminosity of 19.7\fbinv.
Integrated luminosities of 4.8\fbinv (dielectron) and 4.5\fbinv (dimuon) at $\sqrt{s} = 7\TeV$ are used for the double ratio measurements.

Imperfect knowledge of PDFs~\cite{PDF4LHC1,PDF4LHC2} is the dominant source of
theoretical systematic uncertainties in the DY cross section predictions at
low mass. The PDF uncertainty is larger than the achievable
experimental precision, making the double-differential cross section and the double ratio measurements in bins of rapidity an effective
input for PDF constraints.
The inclusion of DY cross section and double ratio data in PDF fits is expected to provide substantial constraints for
the strange quark and the light sea quark PDFs in the small Bjorken $x$ region ($0.001 < x < 0.1$).

The DY differential cross section
has been measured by the CDF, D0, ATLAS, and CMS experiments~\cite{bib:CDFDY1,bib:CDFDY2,bib:D0DY1,bib:ATLAS,bib:ATLASlow,Paper7TeV}.
The current knowledge of the PDFs and the importance of the LHC measurements are reviewed in~\cite{bib:Review,bib:Benchmarking}.
Measuring the DY differential cross section $\rd\sigma/\rd{}m$ is important for various LHC physics analyses.
DY events pose a major source of background
for processes such as top quark pair production, diboson production, and
Higgs measurements with lepton final states,
as well as for searches for new physics beyond the SM, such as the production of high-mass
dilepton resonances.

The differential cross sections are first measured separately for both lepton flavors and found to agree.
The combined cross section measurement is then compared to the
NNLO QCD predictions computed with \FEWZ~3.1~\cite{bib:FEWZ} using the CT10 NNLO PDF.
The $\rd^2\sigma/\rd{}m\,\rd\abs{y}$ measurement is compared to
the NNLO theoretical predictions computed with \FEWZ~3.1 using the CT10 and NNPDF2.1 NNLO PDFs~\cite{CT10, NNPDF}.
\section{CMS detector}

The central feature of the CMS detector is a superconducting solenoid of 6\unit{m}internal diameter and 13\unit{m}length,
providing a magnetic field of 3.8\unit{T}. Within the field volume are a silicon tracker, a crystal electromagnetic calorimeter (ECAL),
and a brass/scintillator hadron calorimeter (HCAL).
The tracker is composed of a pixel detector and a silicon strip tracker, which are used to measure
charged-particle trajectories and cover the full azimuthal angle and the pseudorapidity interval $\abs{\eta} < 2.5$.

Muons are detected with four planes of gas-ionization detectors.
These muon detectors are installed outside the solenoid and
sandwiched between steel layers, which serve both as hadron absorbers and
as a return yoke for the magnetic field flux. They are made using three technologies: drift tubes, cathode
strip chambers, and resistive-plate chambers. Muons are measured in the pseudorapidity window $\abs{\eta} < 2.4$.
Electrons are detected using the energy deposition in the ECAL, which consists of nearly 76\,000 lead tungstate
crystals that are distributed in the barrel region ($\abs{\eta} < 1.479$) and
two endcap ($1.479 < \abs{\eta} < 3$) regions.

The CMS experiment uses a two-level trigger system. The
level-1 trigger, composed of custom processing hardware, selects events
of interest at an output rate of 100\unit{kHz} using information from the calorimeters and muon
detectors~\cite{L1TDR}.  The high-level trigger (HLT) is software based
and further decreases the event collection rate to a few hundred hertz by using the full event
information, including that from the tracker~\cite{HLTTDR}.
A more detailed description of the CMS detector, together with a definition of the coordinate system used
and the relevant kinematic variables, can be found in~\cite{CMS}.
\section{Simulated samples}

Several simulated samples are used
for determining efficiencies, acceptances, and backgrounds from processes that result in two leptons, and for the determination of systematic uncertainties.
The DY signal samples with $\EE$ and $\MM$ final states are generated with the next-to-leading (NLO) generator \POWHEG~\cite{Nason:2004rx,Frixione:2007vw,Alioli:2010xd,Alioli:2008gx} interfaced
with the \PYTHIA v6.4.24~\cite{PYTHIA} parton shower generator.
\PYTHIA is used to model QED final-state radiation (FSR).

The \POWHEG simulated sample is based on NLO calculations, and a correction is applied to take into account
higher-order QCD and electroweak (EW) effects.
The correction factors binned in dilepton rapidity $y$ and transverse momentum $\pt$ are determined in each invariant-mass bin to be the
ratio of the double-differential cross sections calculated at NNLO QCD and NLO EW with \FEWZ~3.1 and at NLO with \POWHEG, as described in~\cite{Paper7TeV}.
The corresponding higher-order effects depend on the dilepton kinematic variables.
Higher-order EW corrections are small in comparison to FSR corrections. They increase for invariant masses in the
\TeVns region~\cite{Li:2012wna}, but are insignificant compared to the experimental
precision for the whole mass range under study.
The NNLO QCD effects are most important in the low-mass region.
The effect of the correction factors on the acceptance
ranges up to 50\% in the low-mass region (below 40\GeV), but is almost negligible in the high-mass region (above 200\GeV).

{\tolerance=800
The main SM background processes are simulated with \POWHEG ($\mathrm{DY}\to \Pgt^{+}\Pgt^{-}$,
single top quark) and with \MADGRAPH~\cite{Madgraph} (\ttbar, diboson events $\PW\PW/\PW\cPZ/\cPZ\cPZ$). Both \POWHEG
and \MADGRAPH are interfaced with the \TAUOLA package~\cite{Tauola}, which handles decays of $\tau$
leptons. The normalization of the \ttbar sample is set to the NNLO cross section of 245.8\unit{pb}~\cite{bib:Txsec}. Multijet QCD
background events are produced with \PYTHIA.
\par}

All generated events are processed through a detailed simulation
of the CMS detector based on \GEANTfour~\cite{Geant4} and are reconstructed using the same algorithms
used for the data. The proton structure is defined using the CT10~\cite{CT10} PDFs.
The simulation includes the effects of multiple interactions per bunch crossing~\cite{CMS_WZ8TeV} (pileup)
with the simulated distribution of the number of interactions per LHC beam crossing corrected to match that observed in data.
\section{Object reconstruction and event selection}
\label{sec:evsel}

The events used in the analysis are selected with a dielectron or a dimuon trigger. Dielectron
events are triggered by the presence of two electron candidates that pass loose requirements
on the electron quality and isolation
with a minimum transverse momentum \pt of 17\GeV for one
of the electrons and 8\GeV for the other. The dimuon trigger requires one muon with $\pt > 17\GeV$ and a second muon
with $\pt > 8\GeV$.

The offline reconstruction of the electrons begins with the clustering of energy depositions in the ECAL.
The energy clusters are then matched to the
electron tracks.
Electrons are identified by means of shower shape variables.
Each electron is required to be consistent with originating from
the primary vertex in the event.
Energetic photons produced in a \PP collision may interact with the detector material
and convert into an electron-positron pair.
The electrons or positrons originating
from such photon conversions are suppressed
by requiring that there be no more than one missing tracker hit
between the primary vertex and the first hit on the reconstructed track
matched to the electron;
candidates are also rejected if they form a pair
with a nearby track that is consistent with a conversion.
Additional details on electron reconstruction
and identification can be found in~\cite{bib:EGM11001,bib:ElectronReco1,bib:ElectronReco2,bib:EEreco3}.
No charge requirements are imposed on the electron pairs
to avoid inefficiency due to nonnegligible charge misidentification.

{\tolerance=800
At the offline muon reconstruction stage, the
data from the muon detectors are matched and fitted to data from the
silicon tracker to form muon candidates.
The muon candidates are required to pass the standard CMS muon identification and track quality criteria~\cite{bib:MUO10004}.
To suppress the background contributions due to muons originating from heavy-quark decays
and nonprompt muons from hadron decays, both muons are required to be isolated from other
particles.
Requirements on the impact parameter and the opening angle between the two muons are further imposed to reject cosmic ray muons.
In order to reject muons from light-meson decays, a common vertex
for the two muons is required.
More details on muon reconstruction and identification can be found in~\cite{Paper7TeV} and~\cite{bib:MUO10004}.
Events are selected for further analysis if they contain oppositely charged muon pairs meeting the above requirements.
The candidate with the highest $\chi^2$ probability from a kinematic fit to the dimuon vertex is selected.
\par}

Electron and muon isolation criteria are based on measuring the sum of energy depositions associated with photons and
charged and neutral hadrons reconstructed and identified by means of the
CMS particle-flow algorithm~\cite{bib:PFold, bib:PF1, bib:PF2, bib:PF3}.
Isolation sums are evaluated in a circular region of the ($\eta$,$\phi$)
plane around the lepton candidate with $\Delta R < 0.3$ (where $\Delta R = \sqrt{\smash[b]{(\Delta\eta)^2+(\Delta\phi)^2}}$),
and are corrected for the contribution from pileup.

Each lepton is required to be within the geometrical acceptance of $\abs{\eta} < 2.4$.
The leading lepton in the event is required to have $\pt > 20\GeV$ and the trailing lepton
$\pt > 10\GeV$, which corresponds to the plateau of the trigger efficiency.
Both lepton candidates in each event used in the offline analysis are required to match HLT trigger objects.

After event selection, the analysis follows a series of steps.
First, backgrounds are estimated. Next, the observed background-subtracted yield is unfolded
to correct for the effects of the migration of events among bins of mass and rapidity due to the detector resolution.
The acceptance and efficiency corrections are then applied.
Finally, the migration of events due to FSR is corrected.
Systematic uncertainties associated with each of the analysis steps are evaluated.
\section{Background estimation}

The major background contributions in the dielectron channel arise from $\Pgt^{+}\Pgt^{-}$ and $\ttbar$ processes
in the low-mass region and from QCD events with multiple jets at high invariant mass.
The background composition is somewhat different in the dimuon final state.
Multijet events and DY production of $\Pgt^{+}\Pgt^{-}$ pairs are the dominant sources of background in the dimuon channel at low invariant mass and
in the region just below the \cPZ\ peak.
Diboson and $\ttbar$ production followed by leptonic decays are the dominant sources of background
at high invariant mass. Lepton pair production in $\gamma\gamma$-initiated processes, where
both initial-state protons radiate a photon, is significant at high mass. The contribution from this
channel is treated as an irreducible background and is estimated
with~\FEWZ~3.1~\cite{bib:PIbkg}.
To correct for this background, a bin-by-bin ratio of the DY cross sections
with and without the photon-induced contribution is calculated. This bin-by-bin correction is
applied after the mass resolution unfolding step,
whereas corrections for other background for which we have simulated events are corrected before.
This background correction is negligible at low mass and in the \cPZ\ peak region, rising to approximately 20\% in the highest mass bin.

In the dielectron channel, the QCD multijet background is estimated with a data sample
collected with the trigger requirement of a single electromagnetic cluster
in the event. Non-QCD events, such as DY, are removed from the data sample using
event selection and event subtraction based on simulation, leaving a sample
of QCD events with characteristics similar to those in the analysis data
sample. This sample is used to estimate the probability for a jet to pass the requirements of the electromagnetic trigger and to be falsely reconstructed as an electron.
This probability is then applied to a sample of events with one electron and one jet to estimate the background contribution from an electron and a jet passing electron
selection requirements. As the contribution from two jets passing the electron selections is
considered twice in the previous method, the contribution from a sample with two jets multiplied
by the square of the probability for jets passing the electron selection requirements is further subtracted.

The QCD multijet background in the dimuon channel is evaluated by selecting a control data sample
before the isolation and charge sign requirements are applied, following the method described in~\cite{bib:CMS_WZ}.

The largest background consists of final states with particles decaying by EW interaction,
producing electron or muon pairs, for example, $\ttbar$, $\Pgt^{+}\Pgt^{-}$, and $\PW\PW$. Notably, these final states
contain electron-muon pairs at twice the rate of electron or muon pairs. These
electron-muon pairs can be cleanly identified from a data sample of $\Pe\Pgm$ events and properly scaled
(taking into account the detector acceptance and efficiency) in order to calculate the background
contribution to the dielectron and dimuon channels.

Background yields estimated from an $\Pe\Pgm$ data sample are used to reduce the systematic uncertainty due to the
limited theoretical knowledge of the cross sections of the SM processes.
The residual differences between background contributions estimated from data and simulation are taken into account in the systematic uncertainty assignment,
as detailed in Section~\ref{sec:syst}.

\begin{figure*}
\centering
\includegraphics[width=0.495\textwidth]{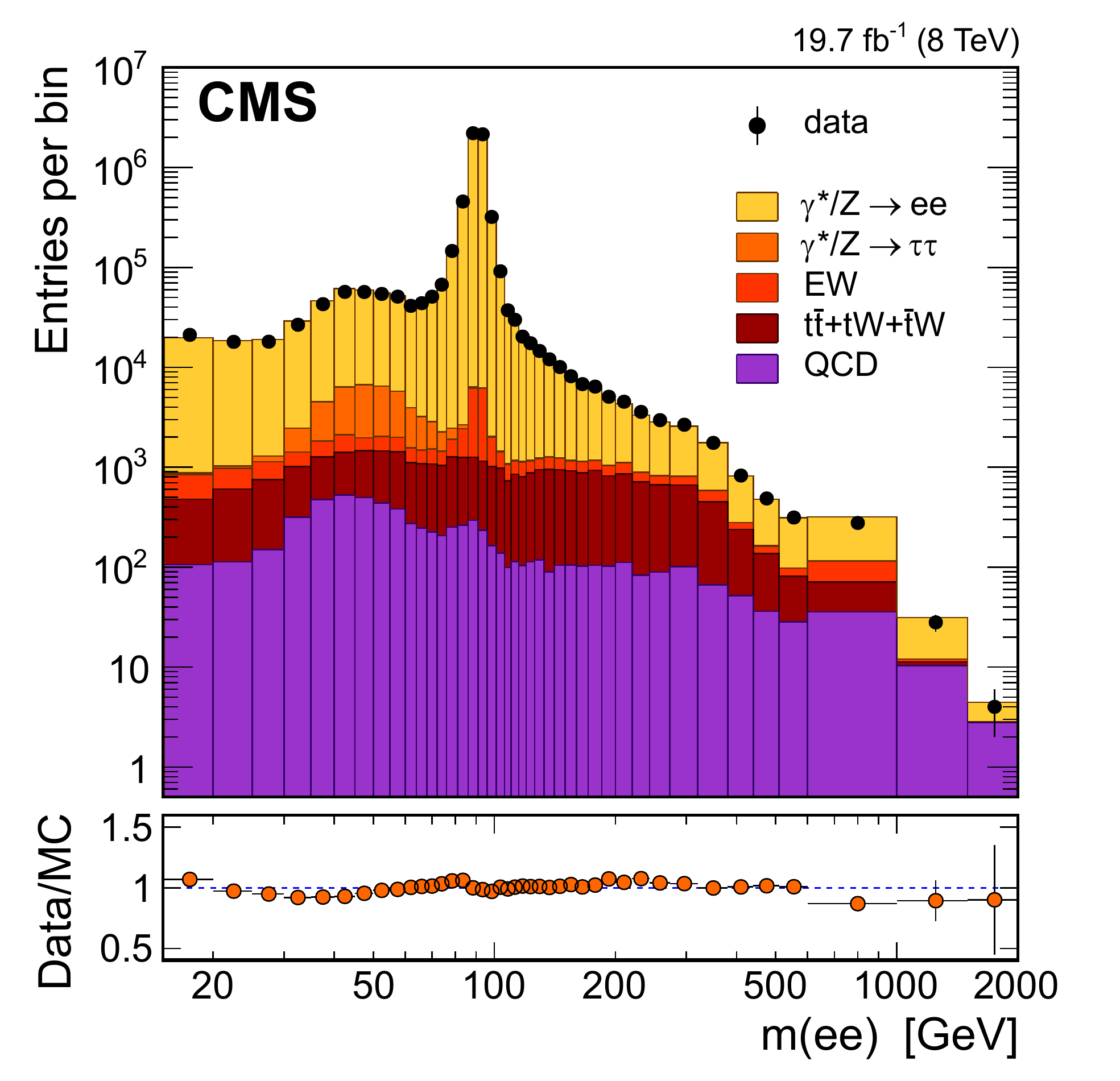}
\includegraphics[width=0.495\textwidth]{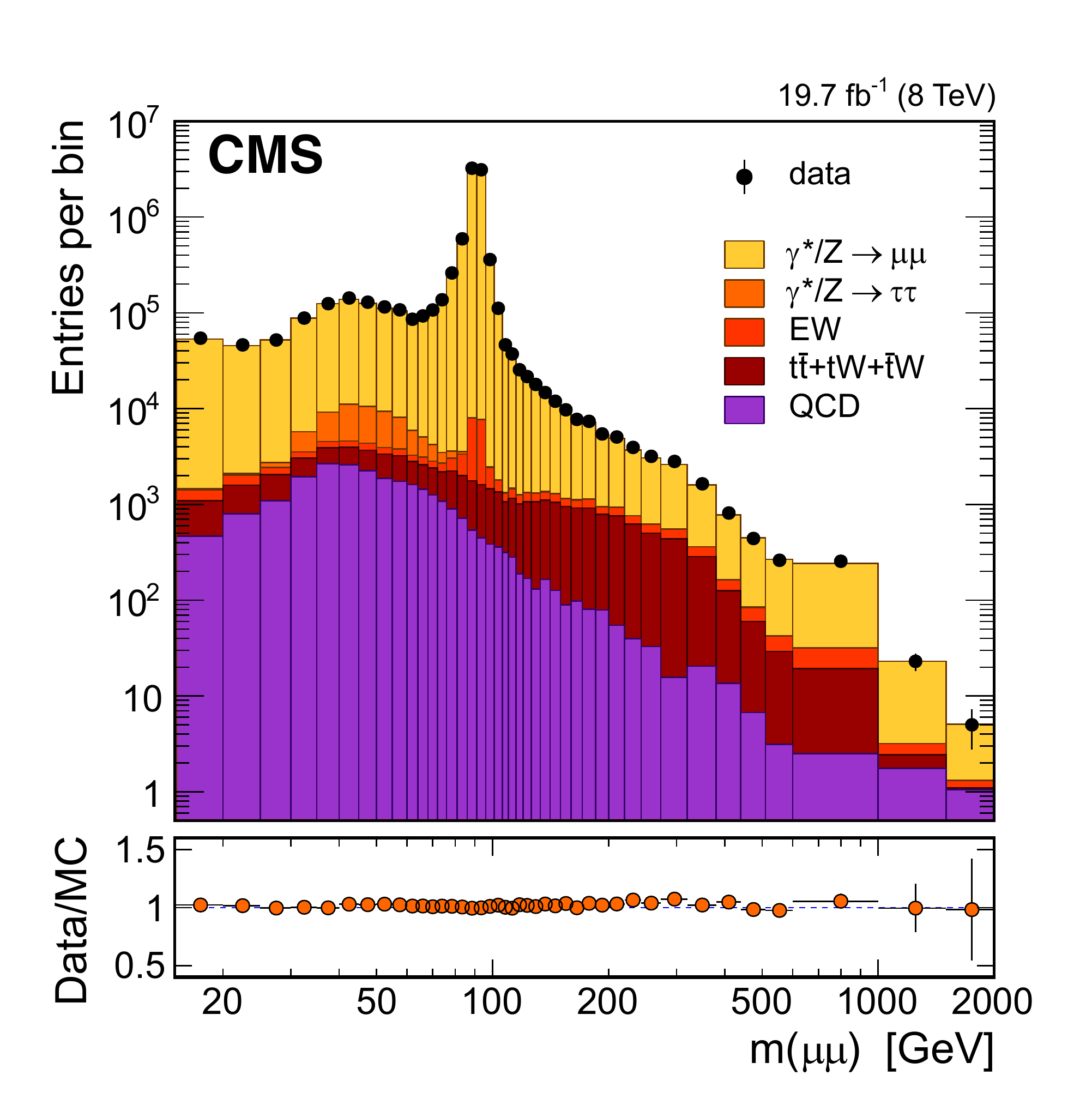}
\caption{
\label{fig:yields1}
The dielectron (left) and dimuon (right) invariant-mass spectra observed in data and predicted by Monte Carlo (MC) simulation
and the corresponding ratios of observed to expected yields.
The QCD multijet contributions in both decay channels
are predicted using control samples in data.
The EW histogram indicates the diboson and $\PW$+jets production.
The simulated signal distributions are based on the NNLO-reweighted \POWHEG sample.
No other corrections are applied.
Error bars are statistical only.
}
\end{figure*}

The dilepton yields for data and simulated events in bins of invariant mass are reported in Fig.~\ref{fig:yields1}.
The photon-induced background is absorbed in the signal distribution so no correction is applied at this stage.
As shown in the figure, the background contribution at low mass is no larger than 5\% in both decay
channels. In the high-mass region, background contamination is more significant, reaching approximately 50\% (30\%)
in the dielectron (dimuon) distribution.
\section{Resolution and scale corrections}

Imperfect lepton energy and momentum measurements can affect
the reconstructed dilepton invariant-mass distributions. Correcting for these effects is important in precise measurements of differential cross sections.

A momentum scale correction to remove a bias in the reconstructed muon momenta
due to the differences in the tracker misalignment between data and simulation and the residual magnetic field mismodeling
is applied following the standard CMS procedure described in~\cite{bib:momcor}.

The electron energy deposits as measured in the ECAL are subject to a set of corrections involving information
both from the ECAL and the tracker, following the standard CMS procedures for the 8\TeV data set~\cite{bib:refHZZ}.
A final electron energy scale correction, which goes beyond the standard set of corrections, is derived from an
analysis of the $\cPZ\to \EE$ peak according to the procedure described in~\cite{bib:CMS_WZ},
and consists of a simple factor of 1.001 applied to the electron energies to account for the different selection used in this analysis.

The detector resolution effects that cause a migration of events among the analysis bins are corrected
through an iterative unfolding procedure~\cite{bib:dagost}. This procedure maps the measured lepton distribution onto the true one,
while taking into account the migration of events in and out of the mass and rapidity range of this measurement.

The effects of the unfolding correction in the differential cross section measurement are approximately
50 (20)\% for dielectron (dimuon) channel in the \cPZ\ peak region, where the invariant-mass spectrum changes steeply.
Less significant effects, of the order of 15\% (5\%) in dielectron (dimuon) channel, are observed in other regions.
The effect on the double-differential cross section measurement is less significant as both the invariant mass
and rapidity bins are significantly wider than the respective detector resolutions.
The effect for dielectrons reaches 15\% in the 45--60\GeV mass region and 5\% at high mass; it is, however, less than 1\% for dimuons over the entire
invariant mass-rapidity range of study.
\section{Acceptance and efficiency}
\label{sec:acceff}

The acceptance $A$ is defined as the fraction of
simulated signal events with both leptons passing the nominal $\pt$ and $\eta$ requirements of the
analysis. It is determined using the NNLO reweighted \POWHEG simulated sample, after the simulation of FSR.

\begin{figure*}
\centering
\includegraphics[width=0.495\textwidth]{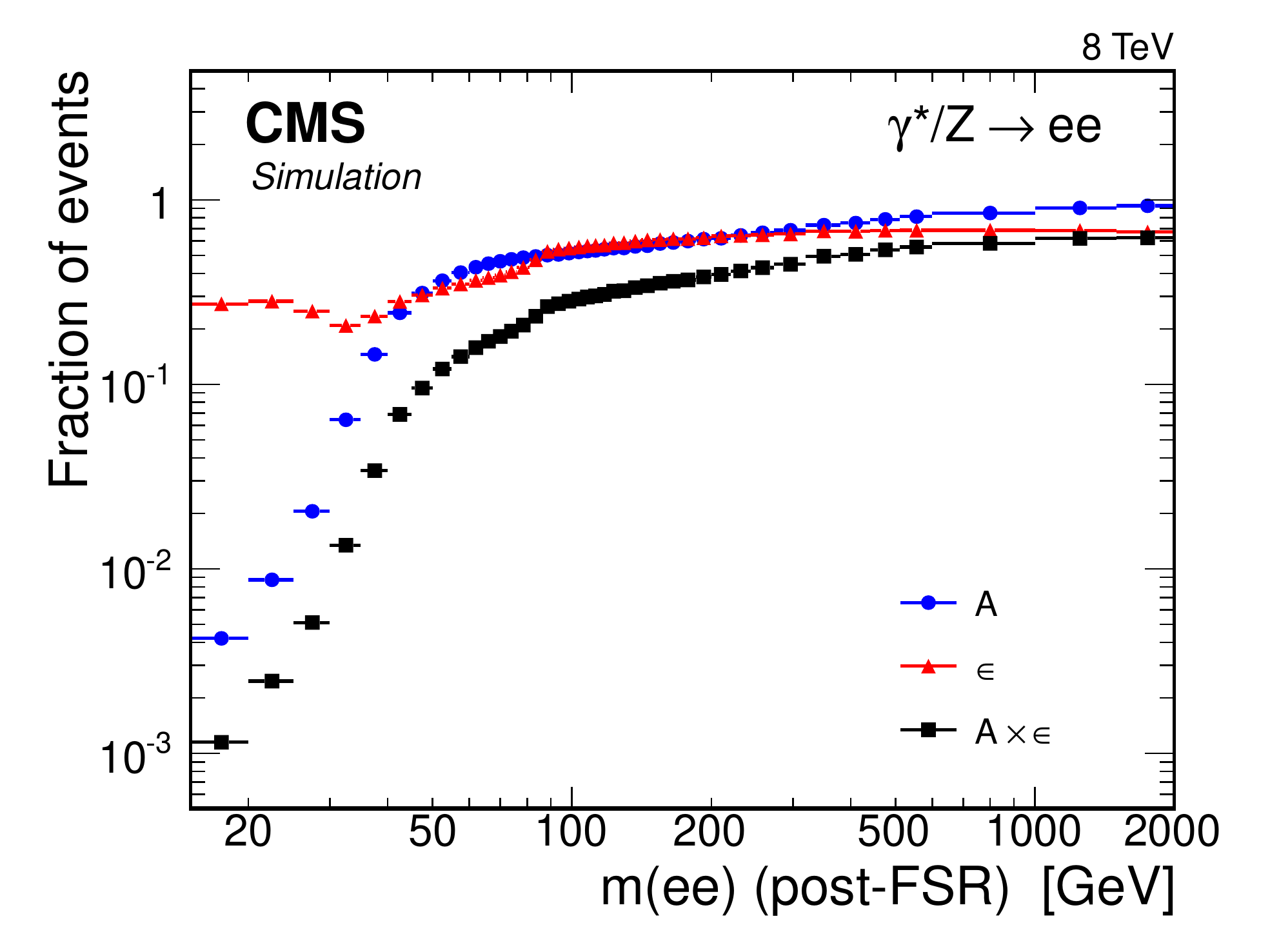}
\includegraphics[width=0.495\textwidth]{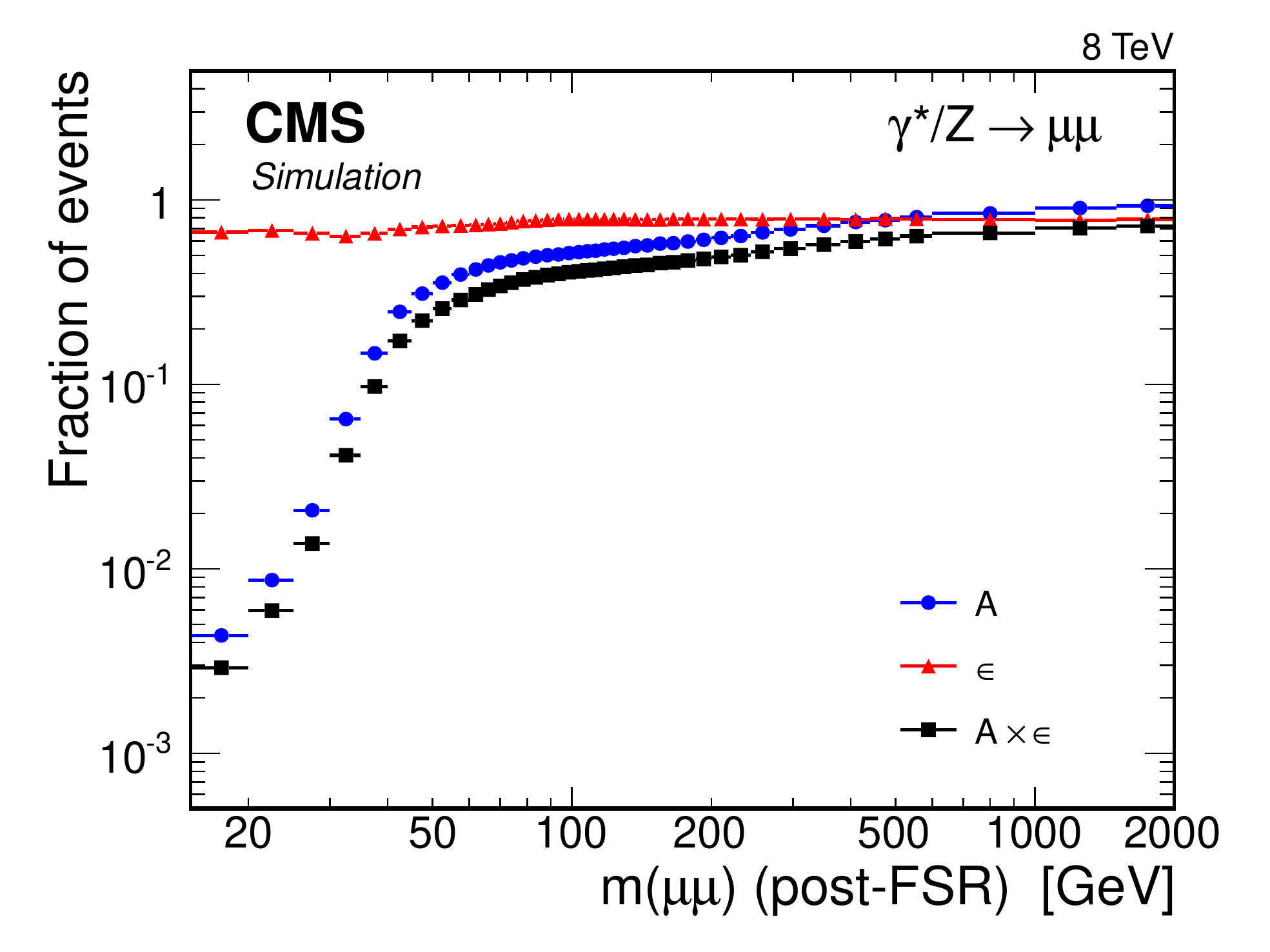}
\caption{
\label{fig:1Dacceff}
The DY acceptance, efficiency, and their product
per invariant-mass bin in the dielectron channel (\cmsLeft) and the dimuon channel (\cmsRight),
where ``post-FSR'' means dilepton invariant mass after the simulation of FSR.
}
\end{figure*}

The efficiency $\epsilon$ is the fraction of events in the DY simulated sample that are inside the acceptance and pass the full selection.
The following equation holds:
\begin{equation}\label{eqn:AccEff}
    A \epsilon \equiv { \frac{N^A}{N^\text{gen}}}
{ \frac{N^{\epsilon}}{N^A} } = {\frac{N^{\epsilon}}{N^\text{gen}}},
\end{equation}
where $N^\text{gen}$ is the number of generated signal events in a given
invariant-mass bin, $N^A$ is the number of events inside the geometrical and
kinematic acceptances, and $N^{\epsilon}$ is the number of events passing
the event selection criteria. Figure~\ref{fig:1Dacceff} shows the acceptance, the efficiency, and their product as functions of the dilepton invariant mass.

The DY acceptance is obtained from simulation. In the lowest mass bin it is only about 0.5\%, rapidly increasing to 50\% in the \cPZ\ peak region and
reaching over 90\% at high mass.

The efficiency is factorized into the reconstruction, identification, and isolation efficiencies and the
event trigger efficiency. The factorization procedure takes into account the asymmetric
\pt selections for the two legs of the dielectron trigger.
The efficiency is obtained from simulation, rescaled with a correction factor that
takes into account differences between data and simulation. The efficiency correction factor is determined
in bins of lepton \pt and $\eta$ using $\cPZ\to \EE (\MM)$ events in data and simulation with the
tag-and-probe method~\cite{bib:CMS_WZ} and is then applied as a weight to simulated events on a per-lepton basis.

A typical dimuon event efficiency is 70--80\%
throughout the entire mass range.
In the dielectron channel, the efficiency at low mass is only 20--40\%
because of tighter lepton identification requirements,
and reaches 65\% at high mass.
The trigger efficiency for events within the geometrical acceptance
is greater than 98\% (93\%) for the dielectron (dimuon) signal.
The efficiency is significantly affected by the pileup in the event.
The effect on the isolation efficiency is up to 5\% (about 1\%)
in the dielectron (dimuon) channel.

A dip in the event efficiency in the mass range 30--40\GeV,
visible in Fig.~\ref{fig:1Dacceff}, is caused by the combination of two factors.
On one hand, the lepton reconstruction and identification efficiencies
decrease as the lepton \pt decreases.
On the other hand, the kinematic acceptance requirements preferentially
select DY events produced beyond the leading order,
which results in higher \pt leptons with higher reconstruction
and identification efficiencies, in the mass range below 30--40\GeV.
The effect is more pronounced for dielectrons than for dimuons
because the electron reconstruction and identification efficiencies
depend more strongly on \pt.

For the dimuon channel the efficiency correction factor is 0.95--1.10, rising up to 1.10 at high dimuon rapidity and falling to 0.95 at low mass.
At low mass, the correction to the muon reconstruction and identification efficiency is dominant, falling to 0.94.
In the dielectron channel, the efficiency correction factor is 0.96--1.05 in the \cPZ~peak region, and
0.90 at low mass. The correction factor rises to 1.05 at high dielectron rapidity.
The correction to the electron identification and isolation efficiency is dominant
in the dielectron channel, reaching 0.93 at low mass and 1.04 at high rapidity.
\section{Final-state QED radiation effects}
\label{sec:FSRsec}

The effect of photon radiation from the final-state leptons (FSR effect) moves the measured invariant mass of the dilepton pair to lower values,
significantly affecting the mass spectrum, particularly in the region below the \cPZ\ peak.
A correction for FSR is performed to facilitate the comparison to the theoretical predictions and to properly combine the measurements in the dielectron and dimuon channels.
The FSR correction is estimated separately from the detector resolution correction
by means of the same unfolding technique.
An additional bin-by-bin correction is applied
for the events in which
the leptons generated before FSR modeling (pre-FSR)
fail the acceptance requirements,
while they pass after the FSR modeling (post-FSR),
following the approach described in~\cite{Paper7TeV}.
The correction for the events not included in the response
matrix is significant at low mass, reaching a maximum of 20\% in the lowest mass bin
and decreasing to negligible levels in the \cPZ\ peak region.

The magnitude of the FSR correction below the \cPZ\ peak is on the order of 40--60\%\,(30--50\%) for the dielectron (dimuon) channel.
In other mass regions, the effect is only 10--15\% in both channels.
In the double-differential cross section measurement, the effect of FSR unfolding is not significant,
typically a few percent, due to a larger mass bin size.

In order to compare the measurements corrected for FSR obtained in analyses with various event generators,
the ``dressed'' lepton quantities can be considered. The dressed lepton four-momentum is defined as
\begin{equation}
\mathbf{p}^{\text{dressed}}_\ell = \mathbf{p}^{\text{post-FSR}}_\ell + \sum \mathbf{p}_{\gamma},
\end{equation}
where all the simulated photons originating from leptons are summed within a cone of $\Delta R < 0.1$.

The correction to the cross sections from the post-FSR to the dressed level reaches a factor of 1.8\,(1.3) in the dielectron\,(dimuon) channel immediately below the \cPZ\ peak;
it is around 0.8 in the low-mass region in both decay channels, and is close to 1.0 at high mass.

\section{Systematic uncertainties}
\label{sec:syst}

\textit{Acceptance uncertainty.}
The dominant uncertainty sources pertaining to the acceptance are (1)~the theoretical uncertainty from
imperfect knowledge of the nonperturbative PDFs contributing to the
hard scattering and (2)~the modeling uncertainty.
The latter comes from the procedure to apply
weights to the NLO simulated sample in order to reproduce NNLO kinematics and
affects mostly the acceptance calculations at very low invariant mass.
The PDF uncertainties for the differential and double-differential cross section measurements
are calculated using the LHAGLUE
interface to the PDF library
LHAPDF 5.8.7~\cite{Bourilkov:2003kk,Whalley:2005nh} by
applying a reweighting technique with asymmetric uncertainties
as described in~\cite{Bourilkov:2006cj}.
These contributions are largest at low and high masses (4--5\%) and
decrease to less than 1\% for masses at the $\cPZ$ peak.

\textit{Efficiency uncertainty.}
The systematic uncertainty in the efficiency estimation consists of two components: the uncertainty in the efficiency correction factor estimation
and the uncertainty related to the number of simulated events.
The efficiency correction factor reflects systematic deviations between data
and simulation. It varies up to 10\% (7\%) for the dielectron (dimuon) channel.
As discussed in Section~\ref{sec:acceff}, single-lepton efficiencies of several types are
measured with the tag-and-probe procedure and are combined into efficiency correction factors.
The tag-and-probe procedure provides the efficiencies for each lepton type and
the associated statistical uncertainties.
A variety of
possible systematic biases in the tag-and-probe procedure have been
taken into account, such as dependence on the binning in single-lepton $\pt$ and
$\eta$, dependence on the assumed shape of signal and background in
the fit model, and the effect of pileup.
In the dielectron channel, this uncertainty is as large as 3.2\% at low mass, and 6\% at
high rapidity in the 200--1500\GeV region.
The uncertainty in the dimuon channel is about 1\% in most of the analysis bins, reaching up to 4\%
at high rapidity in the 200--1500\GeV mass region.
The contribution from the dimuon vertex selection is small because its efficiency correction
factor is consistent with being constant.

\textit{Electron energy scale.} In the dielectron channel, one of the leading systematic uncertainties is associated
with the energy scale corrections for individual electrons. The corrections affect both the
placement of a given candidate in a particular invariant-mass bin and the likelihood of surviving
the kinematic selection. The energy scale corrections are calibrated to a precision of 0.1--0.2\%.
The systematic uncertainties in the measured cross sections
are estimated by varying the electron energy scale by 0.2\%.
The uncertainty is relatively small at low masses. It reaches up to 6.2\%
in the \cPZ\ peak region where the mass bins are the narrowest and the
variation of the cross section with mass is the largest.

\textit{Muon momentum scale.} The uncertainty in the muon momentum scale causes uncertainties
in the efficiency estimation and background subtraction and affects the detector resolution unfolding.
The muon momentum scale is calibrated to 0.02\% precision. The systematic uncertainty in the
measured cross sections is determined by varying the muon momentum scale within its
uncertainty. The largest effect on the final results is observed in the detector
resolution unfolding step, reaching 2\%.

\textit{Detector resolution.}
For both channels, the simulation of the CMS detector, used for
detector resolution unfolding, provides a reliable description of the data.
Possible small systematic errors in the unfolding are related to effects such as differences
in the electron energy scale and muon momentum scale and
uncertainties in FSR simulation and in simulated pileup.
The impact of each of these effects on the measurements is studied
separately, as described in this section.
The detector resolution unfolding procedure itself has been
thoroughly validated, including a variety of closure tests
and comparisons between different event generators;
the systematic uncertainty assigned to the unfolding procedure
is based on the finite size of the simulated samples
and a contribution due to the systematic difference in data and simulation. The latter must be taken into account
because the response matrix is determined from simulation.

\textit{Background uncertainty.} The background estimation uncertainties are evaluated in the same way in both
the dielectron and dimuon channels. The uncertainty in the background is comprised of the
Poissonian statistical uncertainty of predicted backgrounds
and the difference between the predictions from the data and simulation.
The two components are combined in quadrature.
The uncertainty in the background is no larger than 3.0\%\,(1.0\%) at low mass, reaching 16.3\%\,(4.6\%) in the highest mass bin in the dielectron\,(dimuon) channel.

\textit{$\gamma\gamma$-initiated background uncertainty.}
The uncertainty in the correction for $\gamma\gamma$-initiated processes is estimated using \FEWZ~3.1 with the NNPDF2.3QED PDF
and consists of the statistical and PDF uncertainty contributions combined in quadrature.

\textit{FSR simulation.} The systematic uncertainty due to the model-dependent FSR simulation
is estimated using two reweighting techniques described in~\cite{Paper7TeV} with the same procedure in both decay channels.
The systematic uncertainty from modeling the FSR effects is as large as 2.5\%\,(1.1\%) in the dielectron (dimuon) channel
in the 45--60\GeV region.
The systematic uncertainties related to the FSR simulation in the
electron channel primarily affect the detector resolution unfolding
procedure. The impact of these uncertainties is greater for the electron channel
than for the muon channel because of the partial recovery
of FSR photons during the clustering of electron energy in the ECAL.
The effect of the FSR simulation on other analysis steps for the
electron channel is negligible in comparison to other systematic effects
associated with those steps.

\textit{Luminosity uncertainty.}
The uncertainty in the integrated luminosity recorded by CMS
in the 2012 data set is 2.6\%~\cite{bib:LUMI}.

Table~\ref{tab:systele} summarizes the systematic uncertainties for the dielectron and dimuon channels.

\begin{table*}
\centering
\topcaption{
Typical systematic uncertainties (in percent) at low mass (below 40\GeV), in the \cPZ\ peak region ($60 < m < 120$\GeV), and at high mass (above 200\GeV)
for the dielectron and dimuon channels;
``---'' means that the source does not apply.}
\label{tab:systele}
\begin{tabular}{lcc}
Sources & \EE & \MM \\
\hline
Efficiency             & 2.9, 0.5, 0.7 & 1.0, 0.4, 1.8 \\
Detector resolution    & 1.2, 5.4, 1.8 & 0.6, 1.8, 1.6 \\
Background estimation  & 2.2, 0.1, 13.8 & 1.0, 0.1, 4.6 \\
Electron energy scale  & 0.2, 2.4, 2.0 & --- \\
Muon momentum scale    & --- & 0.2, 1.7, 1.6 \\
FSR simulation         & 0.4, 0.3, 0.3 & 0.4, 0.2, 0.5 \\
\hline
Total experimental     & 3.7, 2.5, 14.0 & 1.6, 2.5, 5.4 \\
Theoretical uncertainty& 4.2, 1.6, 5.3 & 4.1, 1.6, 5.3 \\
Luminosity             & 2.6, 2.6, 2.6 & 2.6, 2.6, 2.6\\
\hline
Total                  & 6.3, 6.7, 15.3 & 5.1, 3.9, 8.0 \\
\end{tabular}
\end{table*}

\textit{Systematic uncertainties in the double ratio.}
In the double ratio measurements most of the theoretical
uncertainties are reduced.
The PDF and modeling uncertainties in the acceptance and the systematic uncertainty in the FSR modeling
are fully correlated between 7 and 8\TeV measurements.
The relative uncertainty $\delta\sigma_{s_{i}}/\sigma_{s_{i}}$ in the cross section ratio corresponding to a correlated systematic source of uncertainty $s_{i}$ is
estimated according to
\begin{linenomath}
\begin{equation}
\frac{\delta\sigma_{s_{i}}}{\sigma_{s_{i}}} = \frac{1+\delta_{s_{i}}(8\TeV)}{1+\delta_{s_{i}}(7\TeV)} - 1,
\end{equation}
\end{linenomath}
where the $\delta_{s_{i}}$ are relative uncertainties caused by a source $s_{i}$ in the cross section measurements
at $\sqrt{s} = 7$ and $8\TeV$, respectively.

The systematic uncertainties that are considered uncorrelated between the two center-of-mass energies,
including the uncertainties in efficiency correction estimation, background estimation,
energy scale correction, unfolding, and integrated luminosity, are combined in quadrature.

\section{Results and discussion}

The cross section measurements are first performed separately in the dielectron and dimuon decay channels and then
combined using the procedure described in~\cite{BLUE}.
To assess the sensitivity of the measurement to PDF uncertainties, a comparison to
theoretical calculations is performed using \FEWZ~3.1 with CT10 and NNPDF2.1 NNLO PDFs~\cite{CT10,NNPDF}.
While the theory predictions are presented for NNPDF2.1, similar results are expected from the use of the more recent NNPDF3.0~\cite{Juan2}.

\subsection{Differential cross section \texorpdfstring{$\rd\sigma/\rd{}m$}{d sigma/d m} measurement}

The pre-FSR cross section in the full phase space is calculated as
\begin{equation}
\label{norm}
\sigma^{i} = \frac{N_\mathrm{u}^i}{A^i\epsilon^iL_\text{int}},
\end{equation}
where $N_\mathrm{u}^{i}$ is the number of events after background subtraction and
unfolding procedures for detector resolution and FSR, $A^{i}$ is the acceptance, and
$\epsilon^{i}$ is the efficiency in a given invariant-mass bin $i$; $L_\text{int}$ is the total integrated luminosity.

The cross section in the \cPZ\ peak region is calculated with Eq.~(\ref{norm})
considering the mass region $60 < m < 120\GeV$.

The \cPZ\ peak cross section measurements in the dielectron and dimuon channels are summarized in Table~\ref{tab:normFactor1D_data}.
\begin{table}
\centering
\topcaption{
    Absolute cross section measurements in the \cPZ\ peak region ($60 < m < 120\GeV$).
    The uncertainties in the measurements
    include the experimental and theoretical systematic sources and the uncertainty in the integrated luminosity.
    The statistical component is negligible.}
\label{tab:normFactor1D_data}
\begin{tabular}{ll}
Channel & \multicolumn{1}{c}{Cross section}\\
\hline
Dielectron & $1141 \pm 11\expstat \pm 25\thy \pm 30\lum$\unit{pb}\\
Dimuon & $1135 \pm 11\expstat \pm 25\thy \pm  30\lum$\unit{pb} \\
Combined & $1138 \pm 8\expstat \pm 25\thy \pm 30\lum$\unit{pb}
\end{tabular}
\end{table}
The measurements agree with NNLO theoretical predictions
for the full phase space (\ie, $1137 \pm 36$\unit{pb}, as calculated with \FEWZ~3.1 and CT10 NNLO PDFs),
and also with the previous CMS measurement~\cite{CMS_WZ8TeV}.

The pre-FSR cross section for the full phase space
is calculated in mass bins covering the range 15 to 2000\GeV by means of Eq.~(\ref{norm}).
The results are divided by the invariant-mass bin widths $\Delta m^i$.

The consistency of the differential cross section measurements obtained in the dielectron and
dimuon channels is characterized by a $\chi^2$ probability of 82\%, calculated from the total uncertainties. Therefore the measurements in the two channels are in agreement
and are combined using the procedure defined in~\cite{BLUE}.
Based on the results in the two channels and their
symmetric and positive definite covariance matrices, the estimates of the true cross section values are found as
unbiased linear combinations of the input measurements having a minimum variance~\cite{bib:Lion}.
The uncertainties are considered to be uncorrelated between the two channels,
with the exception of modeling, PDF, and luminosity uncertainties.
The effects of correlations between the analysis bins and different systematic sources
are taken into account in the combination procedure when constructing the covariance matrix.

\begin{figure}
\centering
\includegraphics[width=\cmsFigWidth]{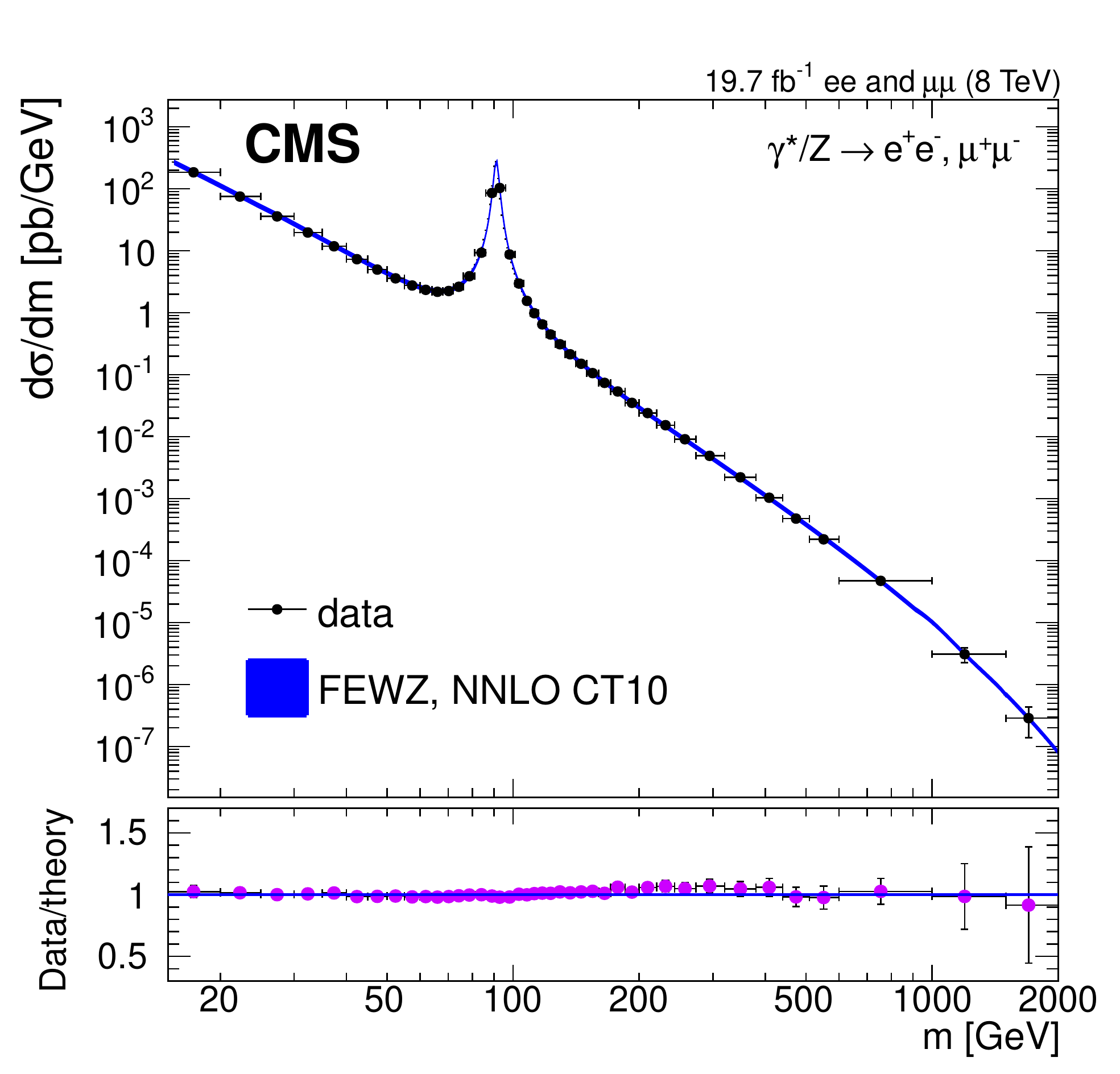}
\caption{
The DY differential cross section as measured in the combined dilepton channel
and as predicted by NNLO \FEWZ~3.1 with CT10 PDF calculations, for the full phase space.
The data point abscissas are computed according to Eq.~(6) in~\cite{bib:pts_plac}.
The $\chi^2$ probability characterizing the consistency of the
predicted and measured cross sections is 91\% with 41 degrees of freedom,
calculated with total uncertainties
while taking into account the correlated errors in the two channels.}
\label{fig:combination}
\end{figure}

The result of the DY cross section measurement in the combined channel is presented in Fig.~\ref{fig:combination}.
The theoretical prediction makes use of the fixed-order NNLO QCD calculation and the NLO EW
correction to DY production initiated by purely weak processes. The $G_{\mu}$ input scheme~\cite{Li:2012wna} is used to
fix the EW parameters in the model. The full spin correlations as well as the
$\gamma^{*}/\cPZ$ interference effects are included in the calculation.
The combined measurement is in agreement
with the NNLO theoretical predictions computed with \FEWZ~3.1 using CT10 NNLO.
The uncertainty band in Fig.~\ref{fig:combination} for the
theoretical calculation represents the combination in quadrature of the statistical uncertainty
from the \FEWZ~3.1 calculation and the 68\% confidence level (CL) uncertainty from the PDFs.
The uncertainties related to QCD evolution scale dependence
are evaluated by varying the renormalization and factorization scales simultaneously between the values 2$m$, $m$, and $m$/2, with $m$ corresponding
to the middle of the invariant mass bin. The scale variation uncertainties reach up to 2\% and
are included in the theoretical error band.

\subsection{Double-differential cross section \texorpdfstring{$\rd^2\sigma/\rd{}m\,\rd\abs{y}$}{d2 sigma/d m abs(y)} measurement}

The pre-FSR cross section in bins of the dilepton invariant mass and the
absolute value of the dilepton rapidity is measured according to
\begin{equation}
\sigma^{ij}_\mathrm{det} = \frac{N_\mathrm{u}^{ij}}{\epsilon^{ij}L_\text{int}}.
\end{equation}
The quantities $N_\mathrm{u}^{ij}$ and $\epsilon^{ij}$
are defined in a given bin $(i,j)$, with $i$ corresponding to the binning
in dilepton invariant mass and $j$ corresponding to the binning in absolute rapidity.
The results are divided by the dilepton absolute rapidity bin widths $\Delta y^{j}$.
The acceptance correction to the full phase space is not applied
to the measurement, in order to keep theoretical
uncertainties to a minimum.

\begin{figure*}
\centering
\includegraphics[width=0.42\textwidth]{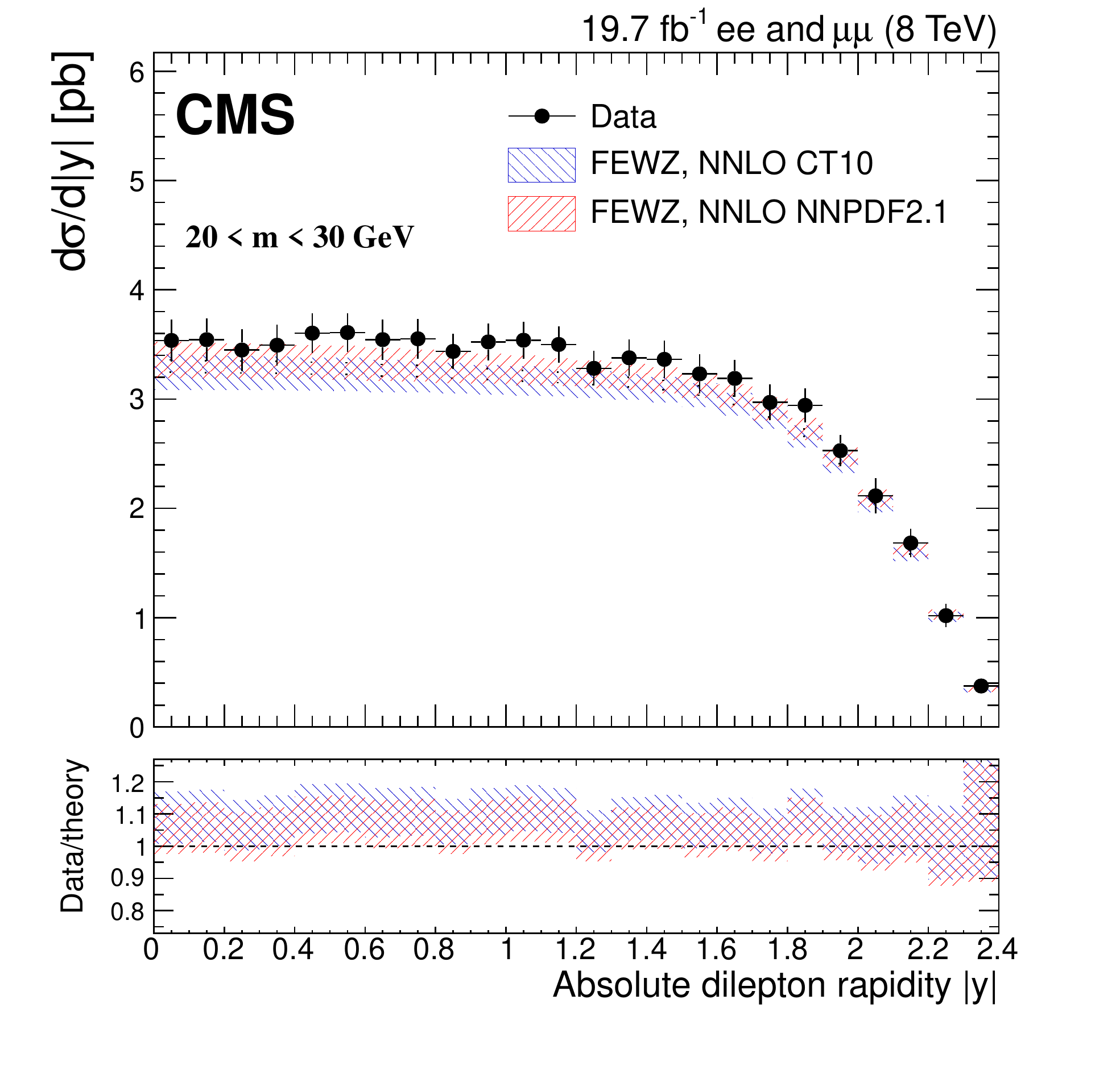}
\includegraphics[width=0.42\textwidth]{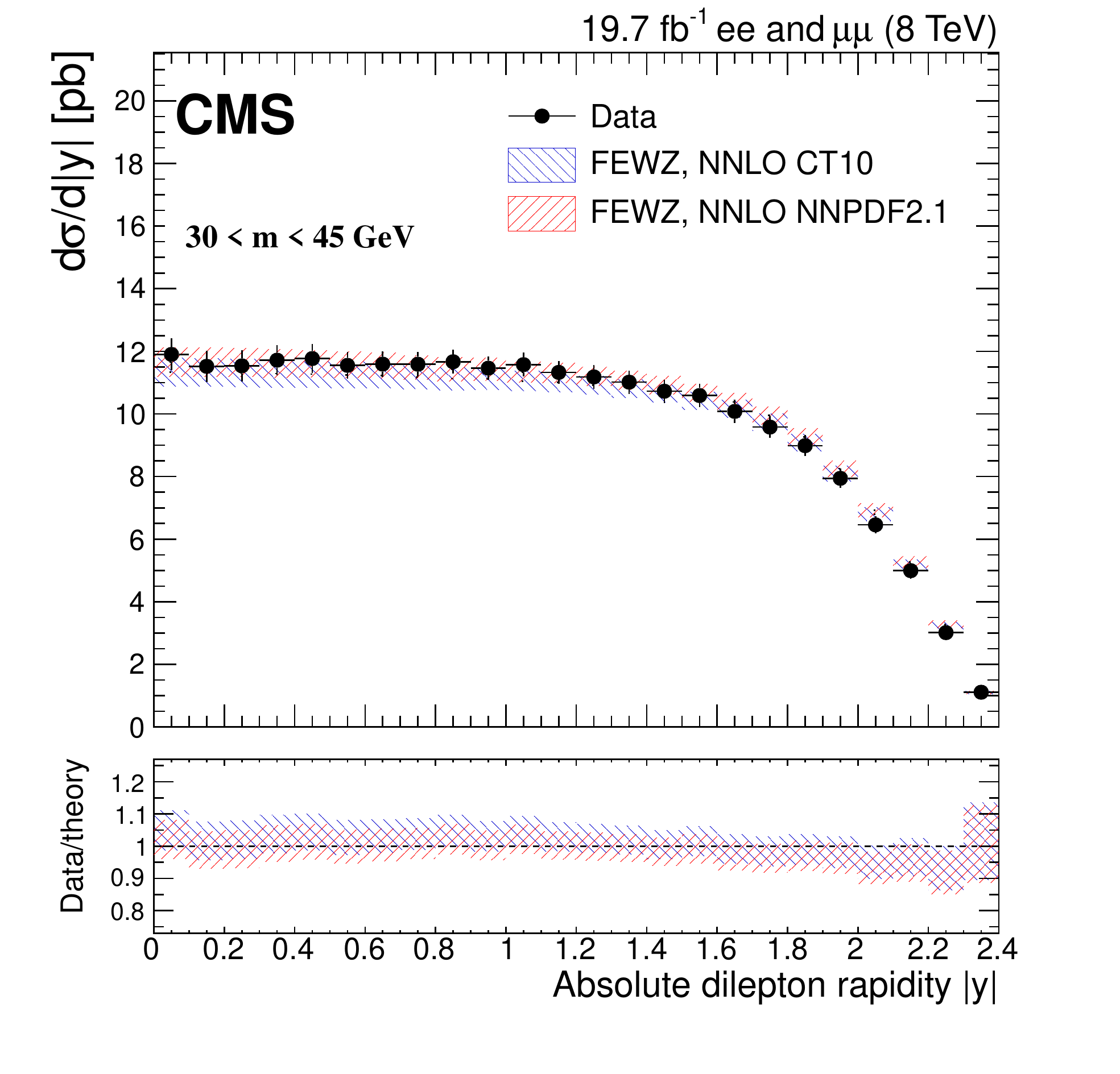}
\includegraphics[width=0.42\textwidth]{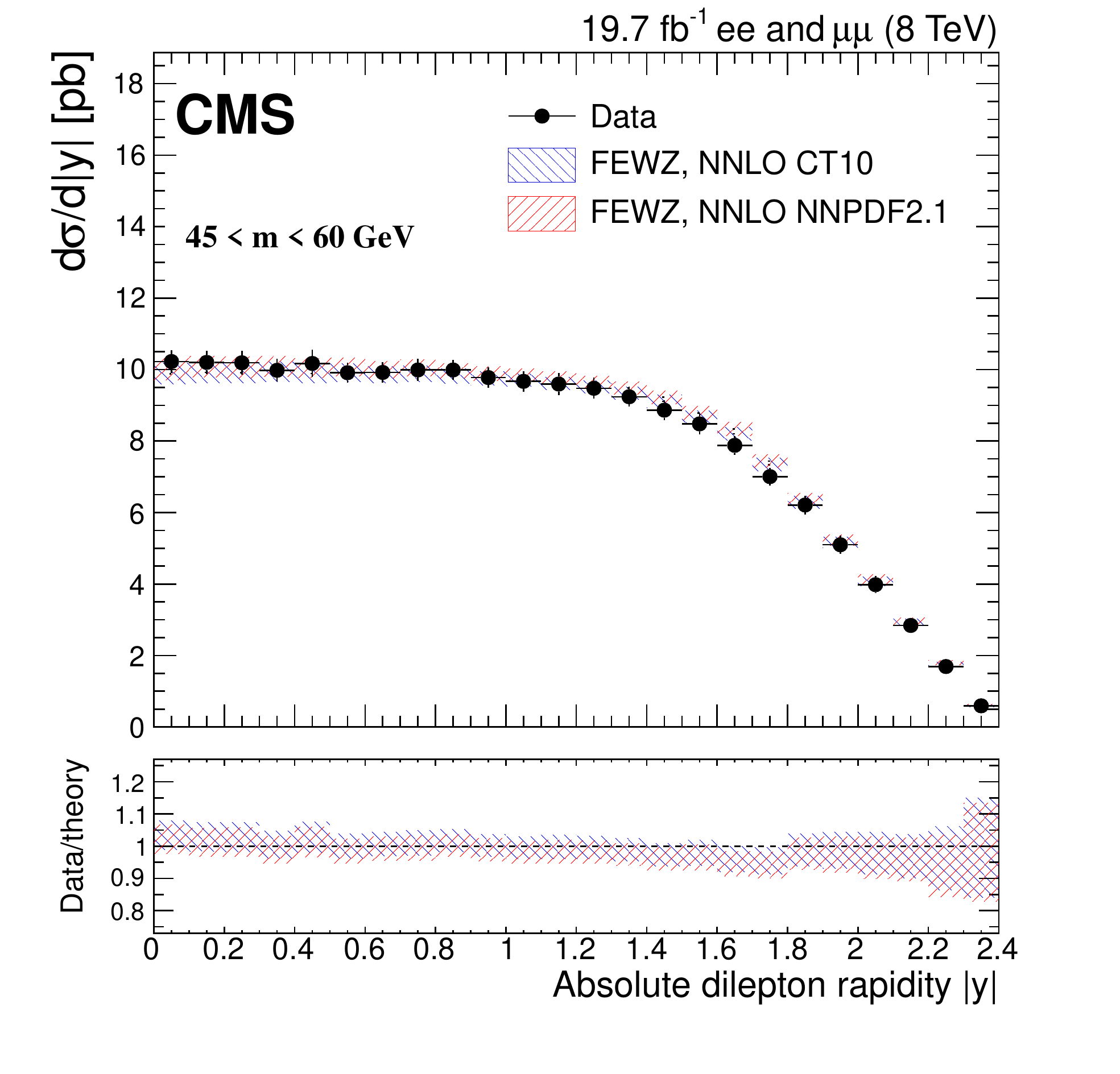}
\includegraphics[width=0.42\textwidth]{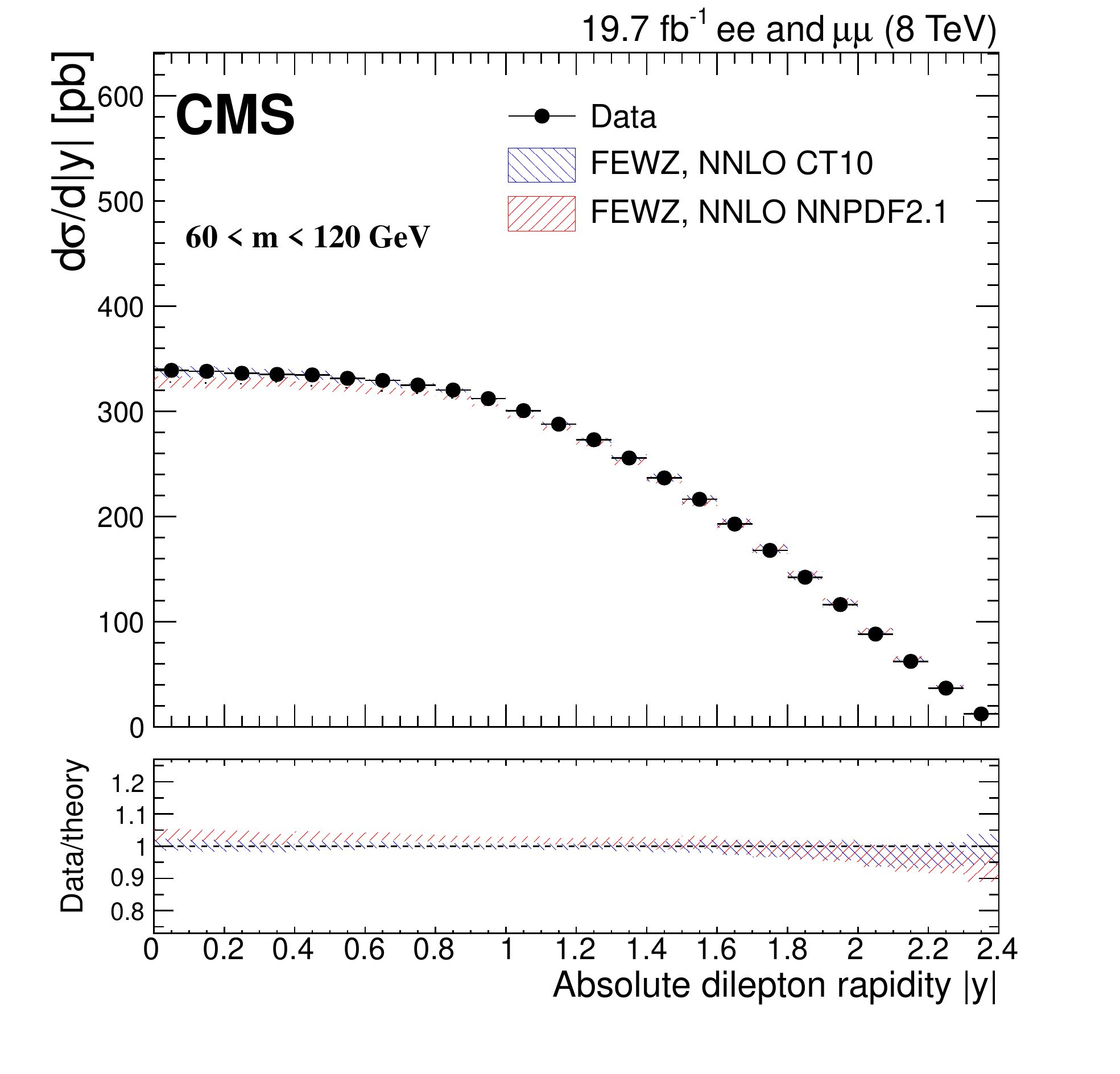}
\includegraphics[width=0.42\textwidth]{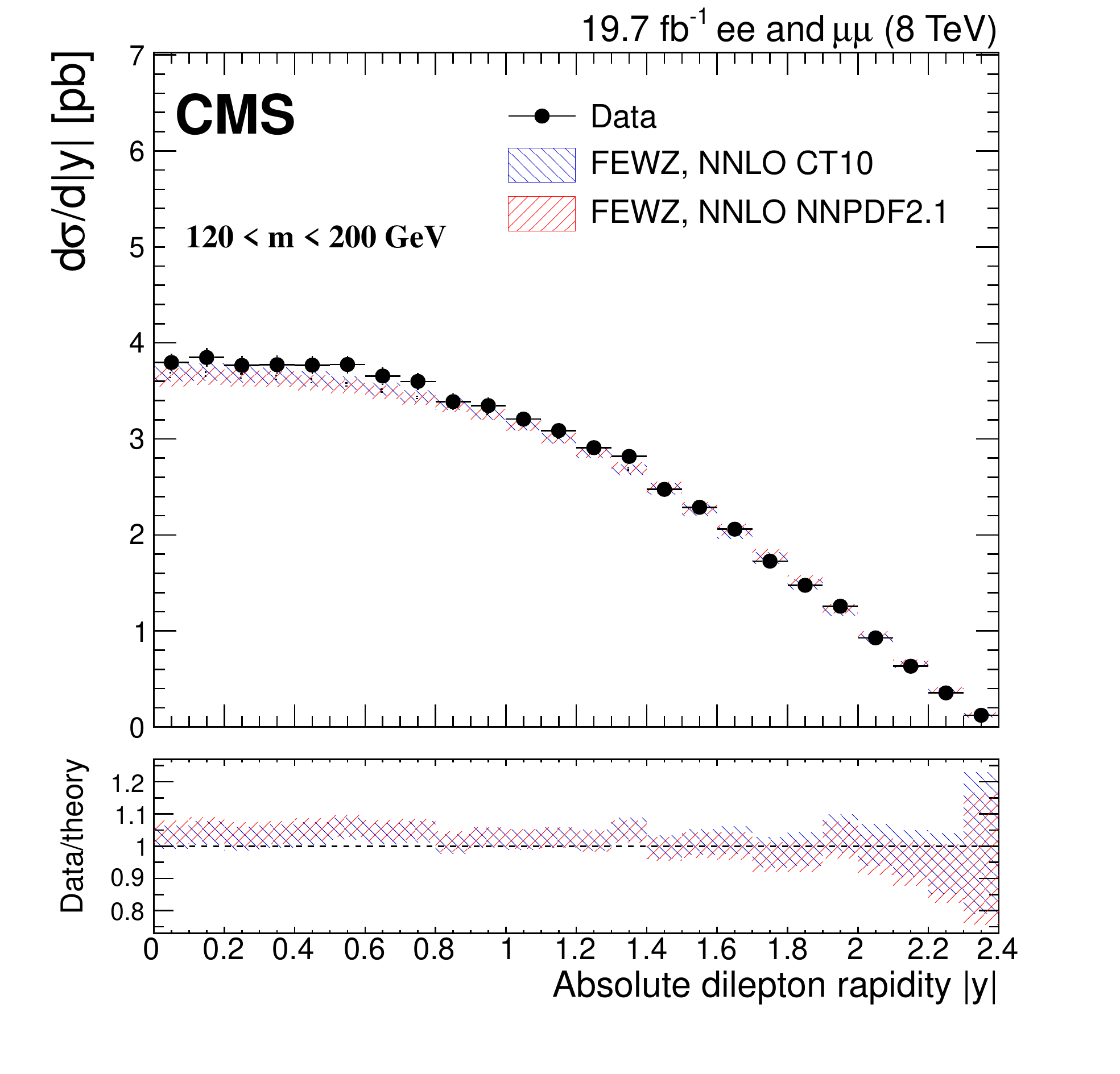}
\includegraphics[width=0.42\textwidth]{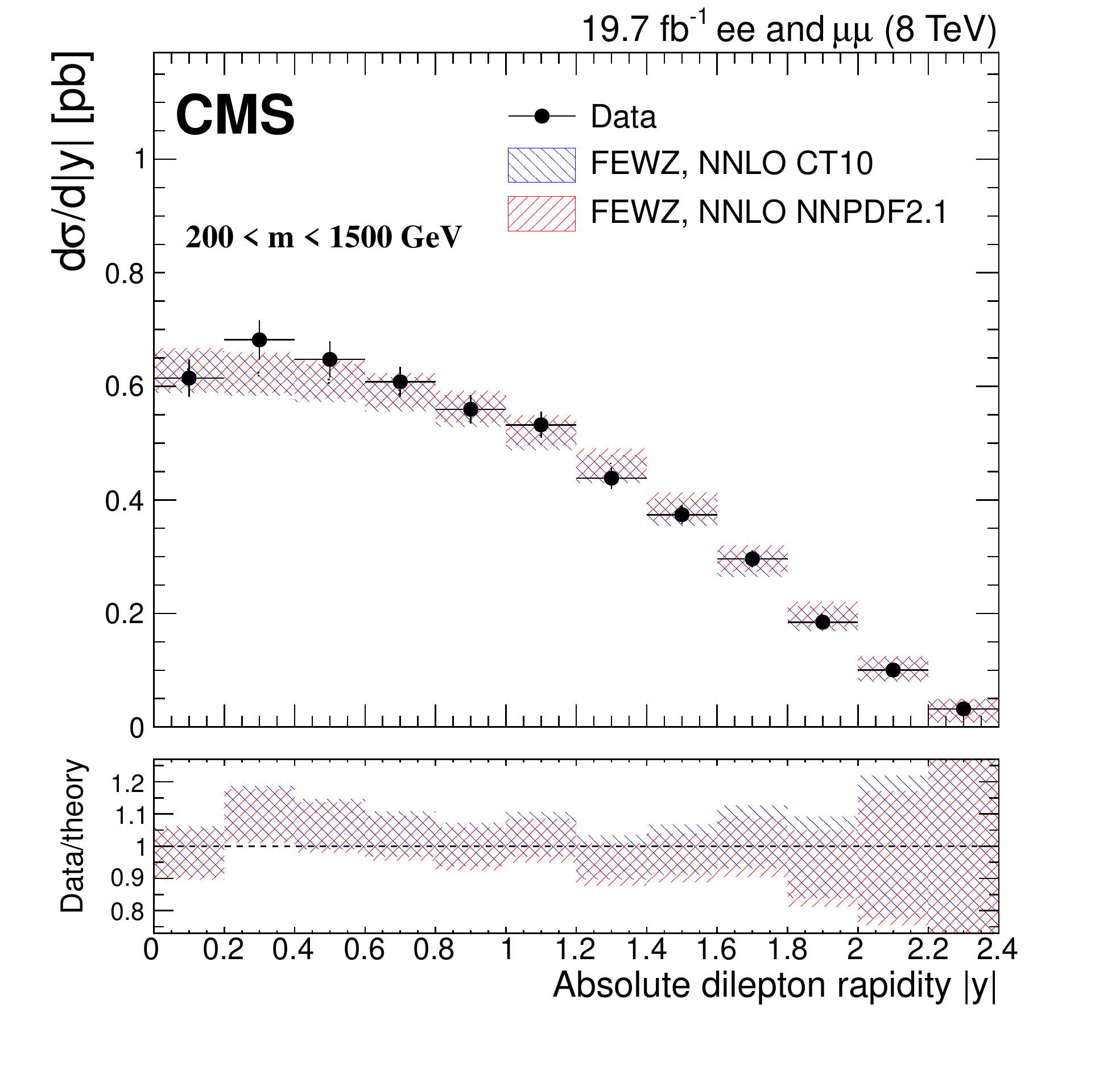}
\caption{
The DY dilepton rapidity distribution $\rd\sigma/\rd\abs{y}$ within the detector acceptance,
plotted for different mass ranges,
as measured in the combined dilepton channel and as predicted by NNLO \FEWZ~3.1 with CT10 PDF and NNLO NNPDF2.1 PDF calculations.
There are six mass bins between 20 and 1500\GeV, from left to right and from top to bottom.
The uncertainty bands in the theoretical predictions
combine the statistical and PDF uncertainties (shaded bands); the latter contributions are dominant.
}
\label{fig:EMu2Drshape}
\end{figure*}

The $\chi^2$ probability characterizing the consistency of
the double-differential cross section measurements in the two channels
is 45\% in the entire invariant mass-rapidity range of study.
The measurements in the two channels are thus in agreement
and are combined using the same procedure as for the differential cross sections described earlier in the section.
Figure~\ref{fig:EMu2Drshape} shows the rapidity distribution $\rd\sigma/\rd\abs{y}$ measured in the
combined dilepton channel with the prediction by \FEWZ~3.1 with the CT10 and NNPDF2.1 NNLO PDF sets.
The cross section is evaluated within the detector acceptance and is plotted for six different mass ranges.

The uncertainty bands in the theoretical expectations include the statistical and the PDF uncertainties
from the \FEWZ~3.1 calculations summed in quadrature.
The statistical uncertainty is significantly smaller than the PDF uncertainty, which
is the dominant uncertainty in the \FEWZ~3.1 calculations.
In general, the PDF uncertainty assignment is different for each PDF set.
The CT10 PDF uncertainties correspond to
90\% CL; to permit a consistent comparison with NNPDF2.1 the uncertainties are scaled to 68\% CL.

In the low-mass region, the results of the measurement are in better agreement with the NNPDF2.1 NNLO than with
the CT10 NNLO estimate, which is systematically lower than NNPDF2.1 NNLO in that region.
The $\chi^2$ probability calculated between data and the theoretical expectation with total uncertainties on the combined results
in the low-mass region is 16\%\,(76\%) for the CT10 (NNPDF2.1) PDFs.
In the \cPZ\ peak region, the two predictions are relatively close to each other and agree well with
the measurements.
The statistical uncertainties in the measurements in the highest mass region are of
the order of the PDF uncertainty.
The corresponding $\chi^2$ probability calculated in the high mass region is 37\%\,(35\%) for the CT10 (NNPDF2.1) PDFs.

\subsection{Double ratio measurements}

{\tolerance=600
The ratios of the normalized differential and double-differential cross sections
for the DY process at the center-of-mass energies of 7 and 8\TeV
in bins of dilepton invariant mass and dilepton absolute rapidity are presented.
The pre-FSR double ratio in bins of invariant mass is calculated following the prescription introduced in ~\cite{bib:DDRojo} according to
\begin{linenomath}
\begin{equation}
\label{simpleR}
R(\PP\to\gamma^{*}/\cPZ\to \ell^{+}\ell^{-}) = \frac{\big(\frac{1}{\sigma_{\cPZ}}\frac{\rd\sigma}{\rd{}m}\big)(8\TeV)}{\big(\frac{1}{\sigma_{\cPZ}}\frac{\rd\sigma}{\rd{}m}\big)(7\TeV)},
\end{equation}
\end{linenomath}
while the pre-FSR double ratio in bins of mass and rapidity is calculated as
\ifthenelse{\boolean{cms@external}}{
\begin{linenomath}
\begin{multline}
\label{simpleRdet}
R_{\mathrm{det}}(\PP\to\gamma^{*}/\cPZ\to \ell^{+}\ell^{-}) = \\ \frac{\big(\frac{1}{\sigma_{\cPZ}}\frac{\rd^2\sigma}{\rd{}m\,\rd\abs{y}}\big)(8\TeV, \pt > 10,\,20\GeV)}{\big(\frac{1}{\sigma_{\cPZ}}\frac{\rd^2\sigma}{\rd{}m\,\rd\abs{y}}\big)(7\TeV, \pt > 9,\,14\GeV)},
\end{multline}
\end{linenomath}
}{
\begin{linenomath}
\begin{equation}
\label{simpleRdet}
R_{\mathrm{det}}(\PP\to\gamma^{*}/\cPZ\to \ell^{+}\ell^{-}) = \frac{\big(\frac{1}{\sigma_{\cPZ}}\frac{\rd^2\sigma}{\rd{}m\,\rd\abs{y}}\big)(8\TeV, \pt > 10,\,20\GeV)}{\big(\frac{1}{\sigma_{\cPZ}}\frac{\rd^2\sigma}{\rd{}m\,\rd\abs{y}}\big)(7\TeV, \pt > 9,\,14\GeV)},
\end{equation}
\end{linenomath}
}
where $\sigma_{\cPZ}$ is the cross section in the \cPZ\ peak region;
$\ell$ denotes $\Pe$ or $\mu$.
The same binning is used for differential measurements at 7 and 8\TeV in order to compute the ratios consistently.
\par}

The double ratio measurements provide a high sensitivity to NNLO QCD effects
and could potentially yield precise constraints on the PDFs;
the theoretical systematic uncertainties in the cross section calculations
at different center-of-mass energies have substantial correlations,
as discussed in Section~\ref{sec:syst}.
Due to cancellation in the double ratio, the effect of the $\gamma\gamma$-initiated processes is
negligible.

\begin{figure}
\centering
\includegraphics[width=\cmsFigWidth]{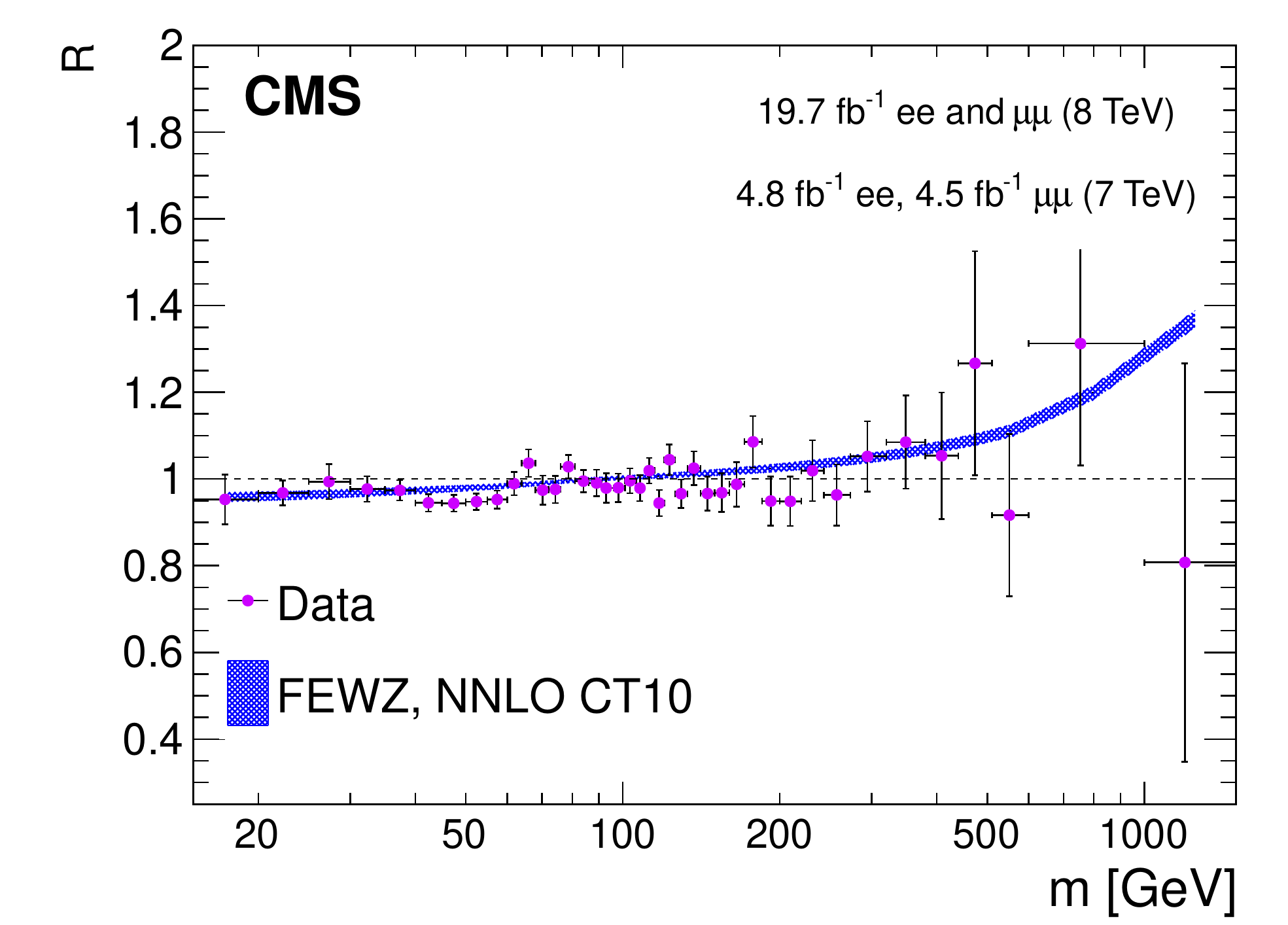}
\caption{
Measured DY double ratios
at center-of-mass energies of 7 and 8\TeV in the combined dilepton channel
as compared to NNLO \FEWZ~3.1 calculations obtained with CT10 NNLO PDF, for the full phase space.
The uncertainty band in the theoretical
predictions combine the statistical and PDF uncertainties; the latter contributions are dominant.
The exact definition of $R$ is given in Eq.~(\ref{simpleR}).
}
\label{fig:1Drshape_double_comb}
\end{figure}

Figure~\ref{fig:1Drshape_double_comb} shows the pre-FSR DY double ratio measurement in the combined (dielectron and dimuon) channel as a function of dilepton invariant mass,
for the full phase space.
The theoretical prediction for the double ratio is calculated using \FEWZ~3.1 with the CT10 NNLO PDF set.
The shape of the distribution is defined entirely by the $\sqrt{s}$ and the Bjorken $x$ dependencies of the PDFs, since the dependence
on the hard scattering cross section is canceled out. In the \cPZ\ peak region, the expected double ratio is close to 1 by definition. It increases linearly as a function
of the logarithm of the invariant mass in the region below 200\GeV, where partons with small Bjorken $x$ contribute the most.
The difference in regions of $x$ probed at 7 and 8\TeV center-of-mass energies
leads to a rapid increase of the double ratio as a function of mass above 200\GeV.

The uncertainty bands in the theoretical prediction of the double ratio include the statistical and the PDF uncertainties
from the \FEWZ~3.1 calculations summed in quadrature.
The experimental systematic uncertainty calculation is described in Section~\ref{sec:syst}.

We observe agreement of the double ratio measurement with the CT10 NNLO PDF theoretical prediction within uncertainties.
The $\chi^2$ probability from a comparison of the
predicted and measured double ratios is 87\% with 40 degrees of freedom,
calculated with the total uncertainties.
At high mass, the statistical component of the uncertainty
becomes significant, primarily from the 7\TeV measurements.

\begin{figure*}
\centering
\includegraphics[width=0.45\textwidth]{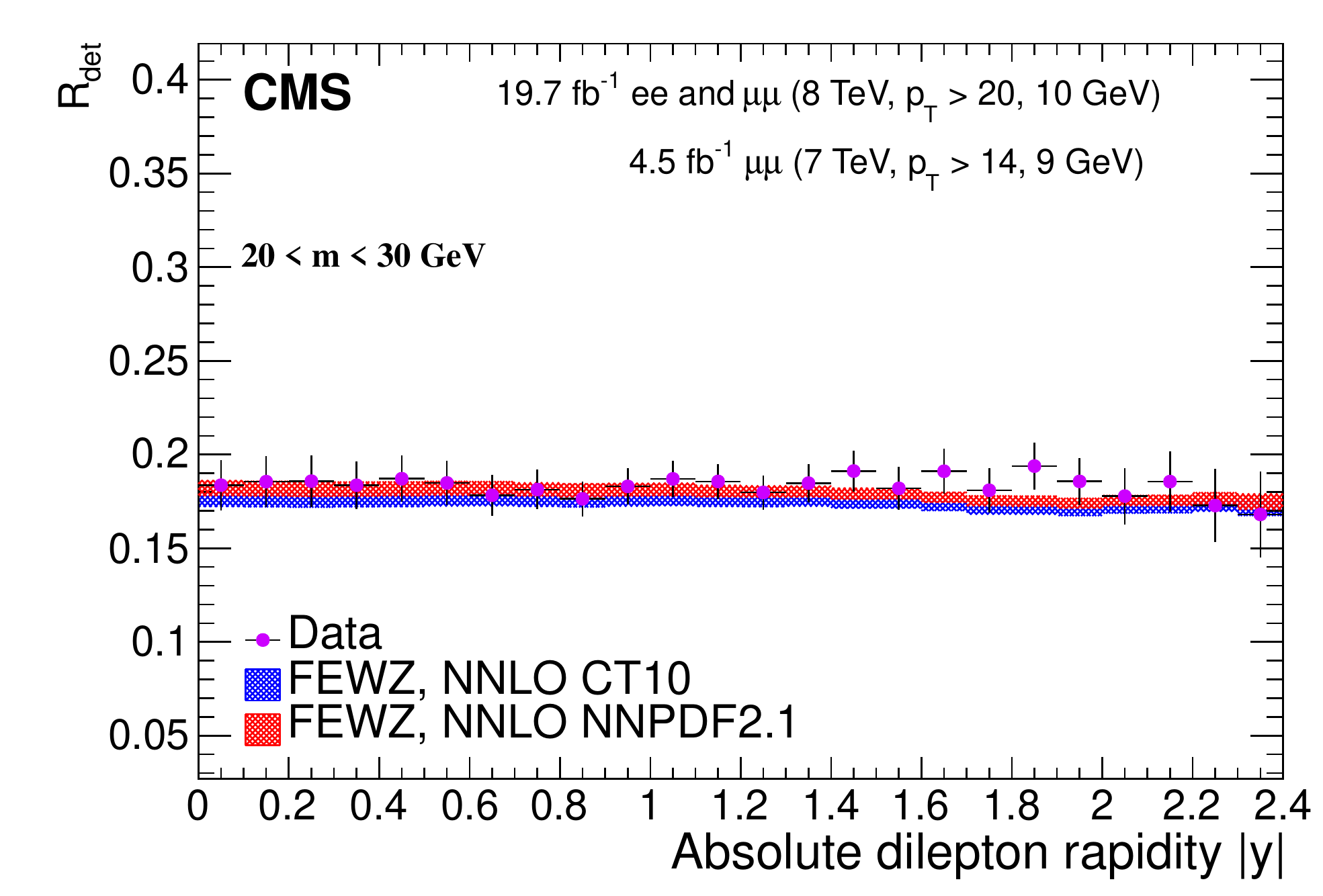}
\includegraphics[width=0.45\textwidth]{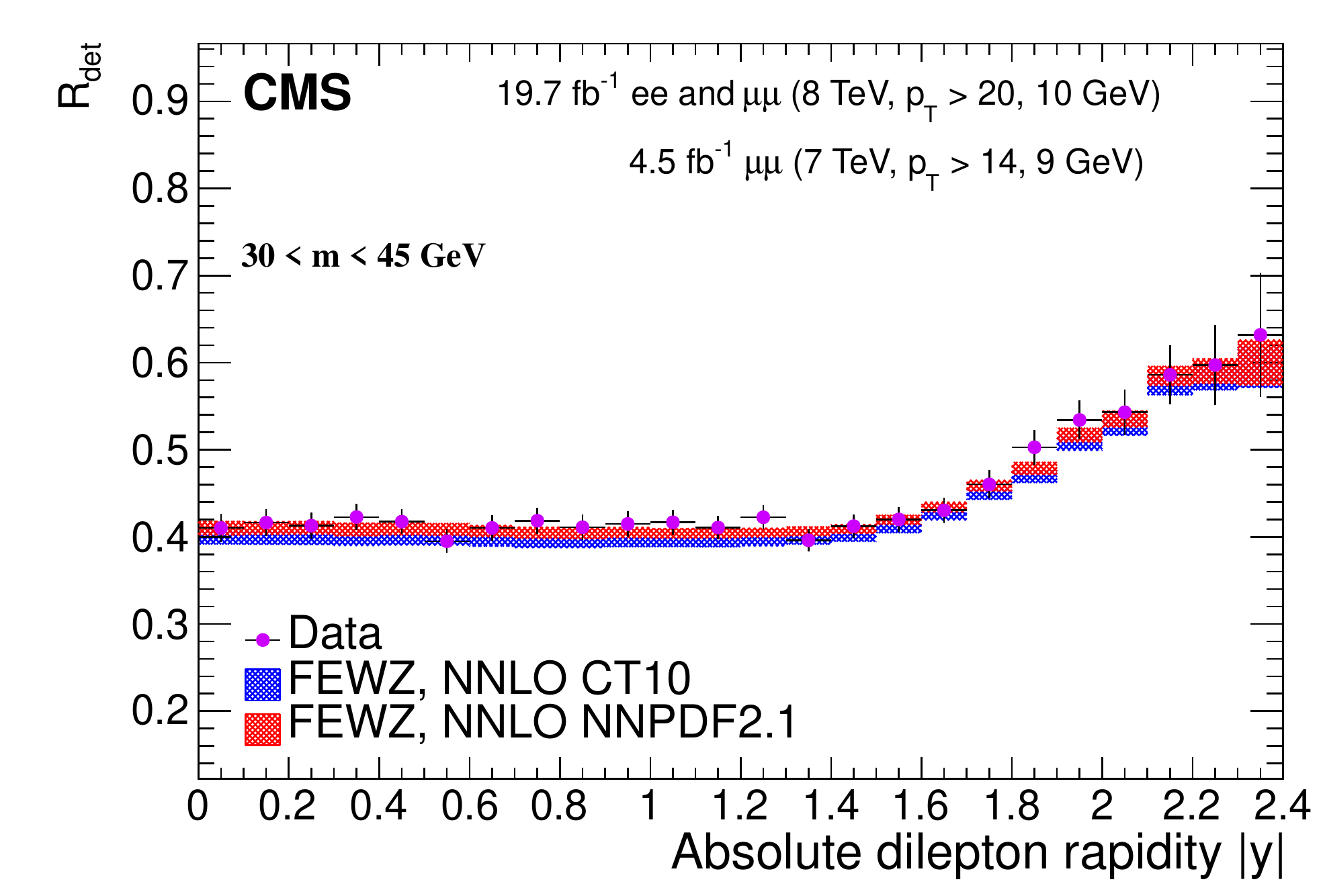}
\includegraphics[width=0.45\textwidth]{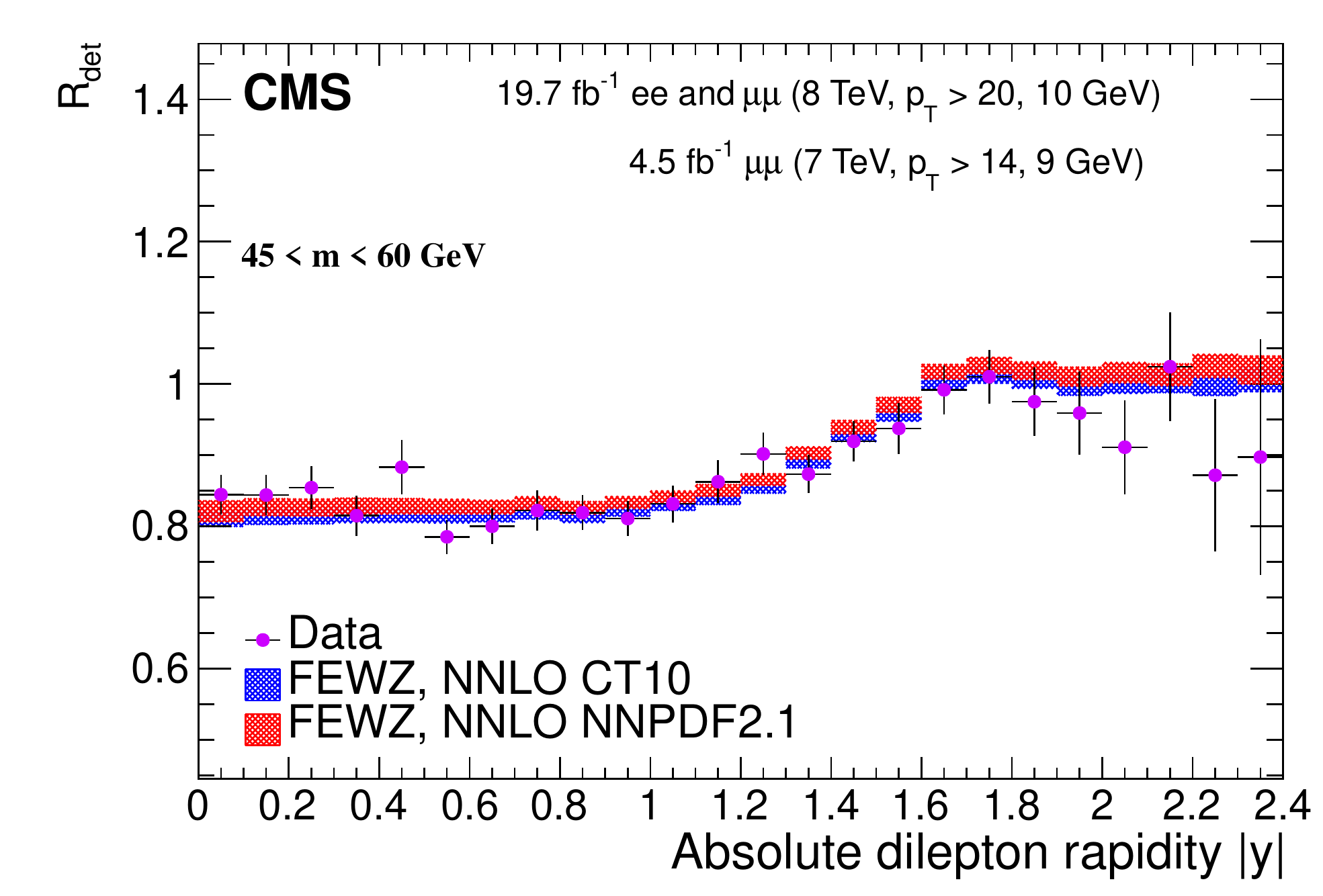}
\includegraphics[width=0.45\textwidth]{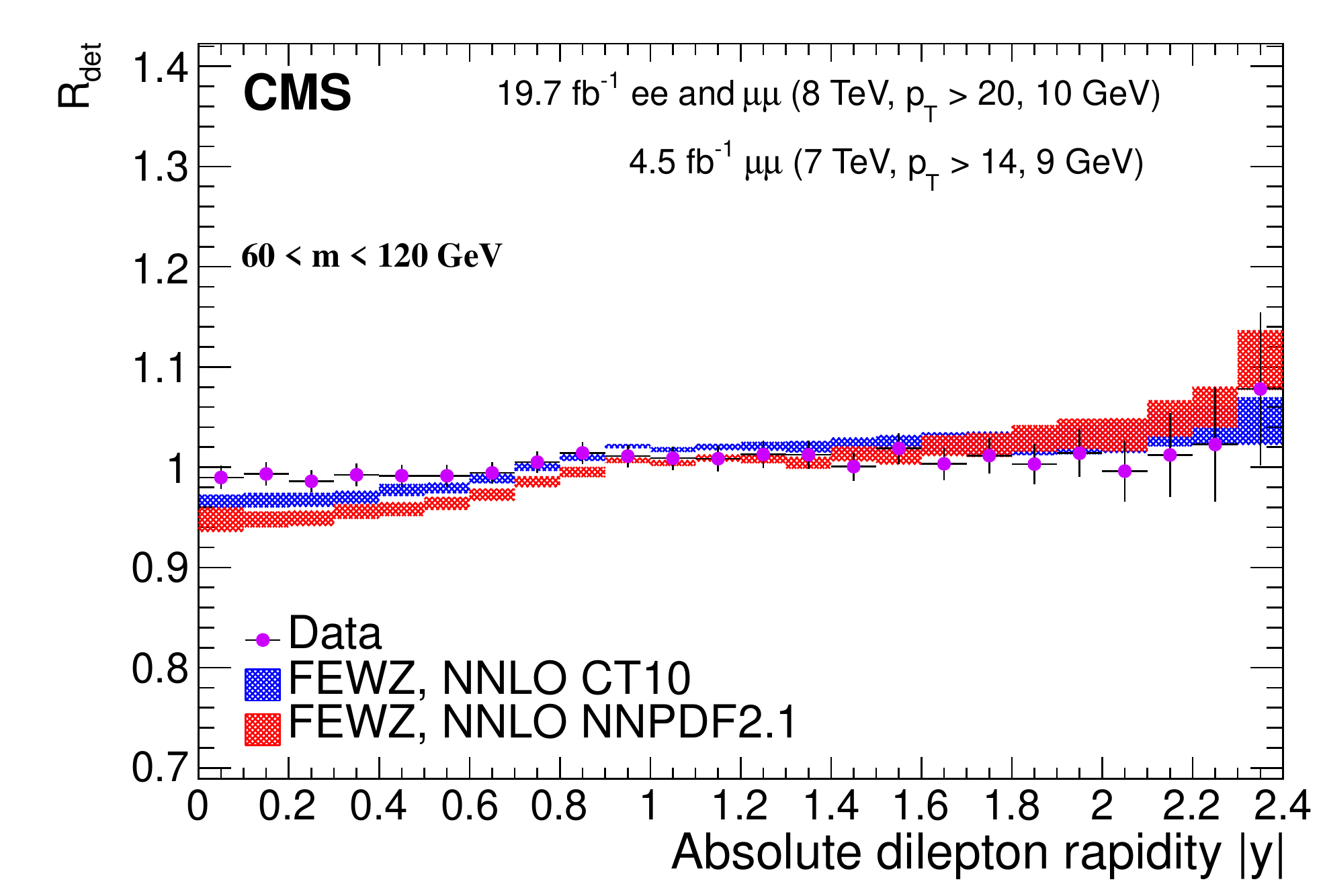}
\includegraphics[width=0.45\textwidth]{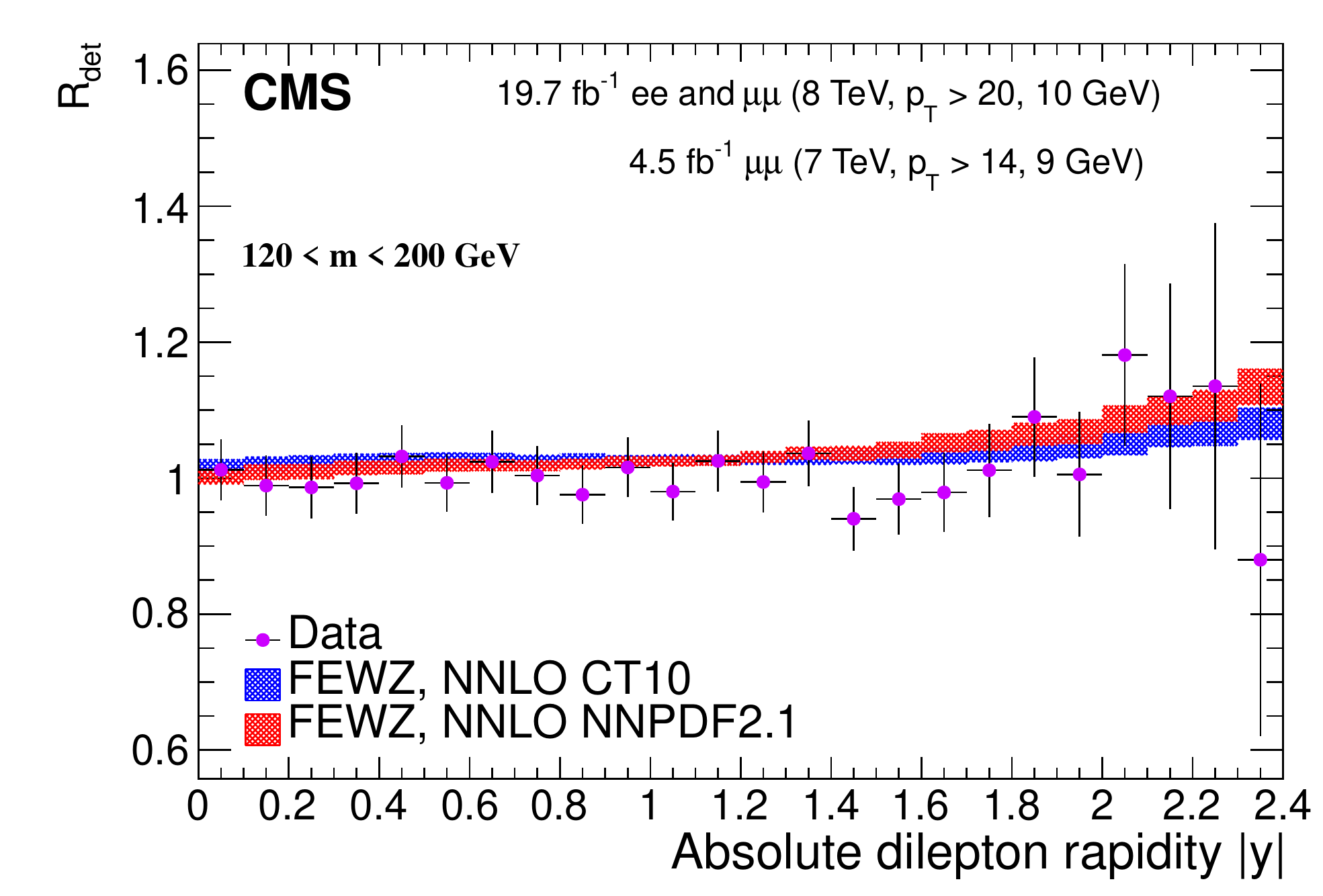}
\includegraphics[width=0.45\textwidth]{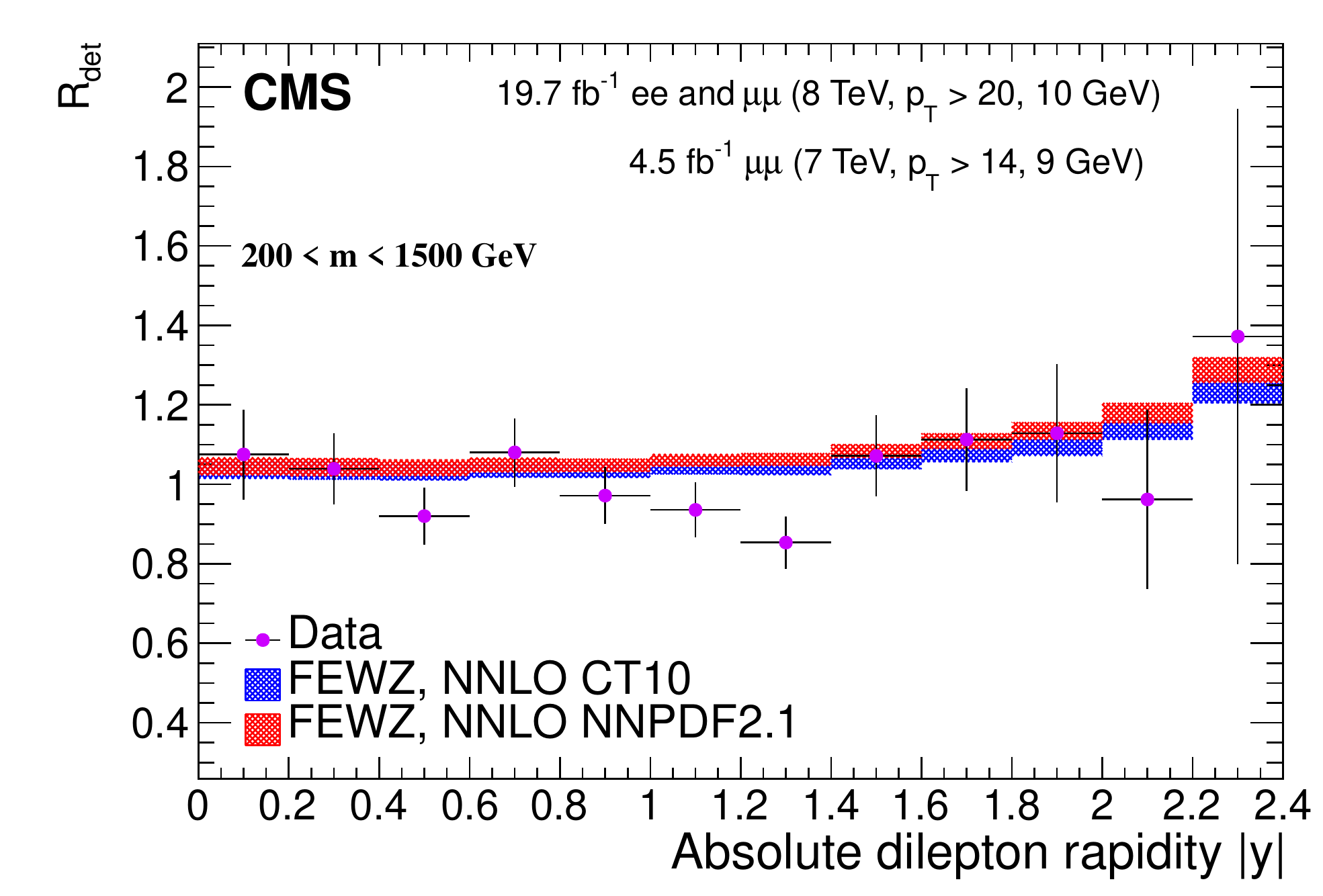}
\caption{
Measured DY double ratios as a function of the absolute dilepton rapidity
within the detector acceptance,
at center-of-mass energies of 7 and 8\TeV,
plotted for different mass ranges
and as
predicted by NNLO \FEWZ~3.1 with CT10 and NNPDF2.1 NNLO PDF calculations.
There are six mass bins between 20 and 1500\GeV, from left to right and from top to bottom.
The uncertainty bands in the theoretical
predictions
combine the statistical and PDF uncertainties (shaded bands); the latter contributions are dominant.
The exact definition of $R_{\mathrm{det}}$ is given in Eq.~(\ref{simpleRdet}).
}
\label{fig:2Drshape_double_comb}
\end{figure*}

The double ratios within the CMS acceptance as measured and as predicted by \FEWZ~3.1 CT10 and NNPDF2.1 NNLO PDF calculations as a function
of dilepton rapidity in six mass bins are summarized in Fig.~\ref{fig:2Drshape_double_comb}.
The measurements having the smallest experimental systematic uncertainty are used in the calculation.
Thus, the 8\TeV measurement entering the numerator is estimated in the combined channel, while the 7\TeV measurement in the denominator is estimated in
the dimuon channel~\cite{Paper7TeV}.
The shape of the theoretical prediction of the double ratio is nearly independent of the dilepton rapidity at
low mass, showing
an increase as a function of rapidity by up to 20\% in the \cPZ\ peak region and at high mass,
and a significant dependence on rapidity in the 30--60\GeV region.
The uncertainty bands in the theoretical predictions of the double ratio include the statistical and the PDF uncertainties
from the \FEWZ 3.1 calculations summed in quadrature.
The uncertainties related to QCD evolution scale dependence
are evaluated by varying the renormalization and factorization scales simultaneously between the values 2$m$, $m$, and $m$/2, with $m$ corresponding
to the middle of the invariant mass bin. The scale variation uncertainties reach up to 2\% and
are included in the theoretical error band.

The double ratio predictions calculated with the CT10 NNLO and NNPDF2.1 NNLO PDFs agree with the measurements.
Below the \cPZ\ peak, NNPDF2.1 NNLO PDF theoretical predictions are in a closer agreement with the measurement. In the \cPZ\ peak region,
a difference in the slope of both theoretical predictions as compared to the measurement is observed in the central absolute rapidity region.
In the high-rapidity and high-mass regions, the effect of the limited number of events in the 7\TeV measurement is significant.
In the 120--200\GeV region, the measurement is at the lower edge of the uncertainty band of the theory predictions.

The DY double-differential cross section and double ratio measurements presented here can be used to impose constraints on the quark
and antiquark PDFs in a wide range of $x$, complementing the data from the fixed-target experiments
with modern collider data.
\section{Summary}

This paper presents measurements of the Drell--Yan differential cross section $\rd\sigma/\rd{}m$ and the
double-differential cross section $\rd^2\sigma/\rd{}m\,\rd\abs{y}$ with proton-proton collision data collected with
the CMS detector at the LHC at a center-of-mass energy of $8\TeV$.  In addition, the first
measurements of the ratios of the normalized differential and double-differential cross sections
for the DY process at center-of-mass energies of 7 and 8\TeV in bins of dilepton invariant mass
and absolute rapidity are presented. A previously published CMS measurement based on 7\TeV
data \cite{Paper7TeV} is used for the double ratio calculations.

The measured inclusive cross section in the \cPZ\ peak region is
$1138 \pm 8\expstat \pm 25\thy \pm  30\lum\unit{pb}$ for the
combination of the dielectron and dimuon channels. This is the most precise measurement of
the cross section in the \cPZ\ peak region at $\sqrt{s} = 8\TeV$ in CMS.
The $\rd\sigma/\rd{}m$ and $\rd^2\sigma/\rd{}m\,\rd\abs{y}$ measurements agree with the NNLO theoretical predictions
computed with \FEWZ~3.1 using the CT10 NNLO and NNPDF2.1 NNLO PDFs.
The double ratio measurement agrees with the theoretical prediction within the systematic and PDF uncertainties.

The experimental uncertainties in the double-differential cross section and the double ratio
measurements presented are relatively small compared to the PDF uncertainties.

\begin{acknowledgments}
\hyphenation{Bundes-ministerium Forschungs-gemeinschaft Forschungs-zentren} We congratulate our colleagues in the CERN accelerator departments for the excellent performance of the LHC and thank the technical and administrative staffs at CERN and at other CMS institutes for their contributions to the success of the CMS effort. In addition, we gratefully acknowledge the computing centers and personnel of the Worldwide LHC Computing Grid for delivering so effectively the computing infrastructure essential to our analyses. Finally, we acknowledge the enduring support for the construction and operation of the LHC and the CMS detector provided by the following funding agencies: the Austrian Federal Ministry of Science, Research and Economy and the Austrian Science Fund; the Belgian Fonds de la Recherche Scientifique, and Fonds voor Wetenschappelijk Onderzoek; the Brazilian Funding Agencies (CNPq, CAPES, FAPERJ, and FAPESP); the Bulgarian Ministry of Education and Science; CERN; the Chinese Academy of Sciences, Ministry of Science and Technology, and National Natural Science Foundation of China; the Colombian Funding Agency (COLCIENCIAS); the Croatian Ministry of Science, Education and Sport, and the Croatian Science Foundation; the Research Promotion Foundation, Cyprus; the Ministry of Education and Research, Estonian Research Council via IUT23-4 and IUT23-6 and European Regional Development Fund, Estonia; the Academy of Finland, Finnish Ministry of Education and Culture, and Helsinki Institute of Physics; the Institut National de Physique Nucl\'eaire et de Physique des Particules~/~CNRS, and Commissariat \`a l'\'Energie Atomique et aux \'Energies Alternatives~/~CEA, France; the Bundesministerium f\"ur Bildung und Forschung, Deutsche Forschungsgemeinschaft, and Helmholtz-Gemeinschaft Deutscher Forschungszentren, Germany; the General Secretariat for Research and Technology, Greece; the National Scientific Research Foundation, and National Innovation Office, Hungary; the Department of Atomic Energy and the Department of Science and Technology, India; the Institute for Studies in Theoretical Physics and Mathematics, Iran; the Science Foundation, Ireland; the Istituto Nazionale di Fisica Nucleare, Italy; the Ministry of Science, ICT and Future Planning, and National Research Foundation (NRF), Republic of Korea; the Lithuanian Academy of Sciences; the Ministry of Education, and University of Malaya (Malaysia); the Mexican Funding Agencies (CINVESTAV, CONACYT, SEP, and UASLP-FAI); the Ministry of Business, Innovation and Employment, New Zealand; the Pakistan Atomic Energy Commission; the Ministry of Science and Higher Education and the National Science Centre, Poland; the Funda\c{c}\~ao para a Ci\^encia e a Tecnologia, Portugal; JINR, Dubna; the Ministry of Education and Science of the Russian Federation, the Federal Agency of Atomic Energy of the Russian Federation, Russian Academy of Sciences, and the Russian Foundation for Basic Research; the Ministry of Education, Science and Technological Development of Serbia; the Secretar\'{\i}a de Estado de Investigaci\'on, Desarrollo e Innovaci\'on and Programa Consolider-Ingenio 2010, Spain; the Swiss Funding Agencies (ETH Board, ETH Zurich, PSI, SNF, UniZH, Canton Zurich, and SER); the Ministry of Science and Technology, Taipei; the Thailand Center of Excellence in Physics, the Institute for the Promotion of Teaching Science and Technology of Thailand, Special Task Force for Activating Research and the National Science and Technology Development Agency of Thailand; the Scientific and Technical Research Council of Turkey, and Turkish Atomic Energy Authority; the National Academy of Sciences of Ukraine, and State Fund for Fundamental Researches, Ukraine; the Science and Technology Facilities Council, UK; the US Department of Energy, and the US National Science Foundation.

Individuals have received support from the Marie-Curie programme and the European Research Council and EPLANET (European Union); the Leventis Foundation; the A. P. Sloan Foundation; the Alexander von Humboldt Foundation; the Belgian Federal Science Policy Office; the Fonds pour la Formation \`a la Recherche dans l'Industrie et dans l'Agriculture (FRIA-Belgium); the Agentschap voor Innovatie door Wetenschap en Technologie (IWT-Belgium); the Ministry of Education, Youth and Sports (MEYS) of the Czech Republic; the Council of Science and Industrial Research, India; the HOMING PLUS programme of Foundation for Polish Science, cofinanced from European Union, Regional Development Fund; the Compagnia di San Paolo (Torino); the Consorzio per la Fisica (Trieste); MIUR project 20108T4XTM (Italy); the Thalis and Aristeia programmes cofinanced by EU-ESF and the Greek NSRF; and the National Priorities Research Program by Qatar National Research Fund.
\end{acknowledgments}
\bibliography{auto_generated}
\cleardoublepage \appendix\section{The CMS Collaboration \label{app:collab}}\begin{sloppypar}\hyphenpenalty=5000\widowpenalty=500\clubpenalty=5000\textbf{Yerevan Physics Institute,  Yerevan,  Armenia}\\*[0pt]
V.~Khachatryan, A.M.~Sirunyan, A.~Tumasyan
\vskip\cmsinstskip
\textbf{Institut f\"{u}r Hochenergiephysik der OeAW,  Wien,  Austria}\\*[0pt]
W.~Adam, T.~Bergauer, M.~Dragicevic, J.~Er\"{o}, M.~Friedl, R.~Fr\"{u}hwirth\cmsAuthorMark{1}, V.M.~Ghete, C.~Hartl, N.~H\"{o}rmann, J.~Hrubec, M.~Jeitler\cmsAuthorMark{1}, W.~Kiesenhofer, V.~Kn\"{u}nz, M.~Krammer\cmsAuthorMark{1}, I.~Kr\"{a}tschmer, D.~Liko, I.~Mikulec, D.~Rabady\cmsAuthorMark{2}, B.~Rahbaran, H.~Rohringer, R.~Sch\"{o}fbeck, J.~Strauss, W.~Treberer-Treberspurg, W.~Waltenberger, C.-E.~Wulz\cmsAuthorMark{1}
% ...
\vskip\cmsinstskip
% ...
1:~~Also at Vienna University of Technology, Vienna, Austria\\
2:~~Also at CERN, European Organization for Nuclear Research, Geneva, Switzerland\\
% ...

\end{sloppypar}
\end{document}